# Thermodynamic Modeling of Pure Elements from 0 K with Uncertainty Quantification using PyCalphad and ESPEI

Alexander M. Richter[1*], Abdulmonem Obaied[2,3], Irina Roslyakova[2,3], Boris Wilthan[4], Allison M. Beese[1], Zi-Kui Liu[1]

[1] Pennsylvania State University, Department of Materials Science and Engineering, University Park, PA 16802, USA

[2] Ruhr-Universität Bochum, ICAMS, Universitätsstrasse 150, Bochum 44801, Germany

[3] GTT-Technologies, Kaiserstraße 103, Herzogenrath 52134, Germany

[4] National Institute of Standards and Technology, Boulder, Colorado 80305, USA

*e-mail address of corresponding author: amr8004@psu.edu

**Abstract**

Thermodynamic modeling of pure elements is the foundation of the CALPHAD modeling of engineering materials. Recently, multiple physics-based models have been proposed to describe Gibbs energy of pure elements down to 0 K, extending from 298.15 K in the current CALPHAD modeling. To enable their systematic and quantitative comparison and adoption, those thermodynamic models of pure elements are implemented into the open-source software packages PyCalphad and ESPEI in the present work for evaluation of model parameters and model fitness. PyCalphad and ESPEI are suitable tools for implementation of these models for high throughput CALPHAD modeling of multicomponent materials. Particularly, Markov Chain Monte Carlo used in ESPEI allows for uncertainty quantification of model parameters and model predictions. Through the remodeling of 41 pure elements, the present work demonstrates the quantitative comparison of modeling of pure elements with different models and enables the efficient development of multicomponent systems with continuously improved CALPHAD description of pure elements.

# 1 Introduction

Thermodynamic modeling based on the CALPHAD approach[1], [2], [3], [4] is the backbone of the Integrated Computational Materials Engineering (ICME)[5], [6] and the US Materials Genome Initiative[7] . While CALPHAD is typically referred to as calculation of phase diagrams pioneered by Kaufman[1], [8] , it covers both equilibrium and nonequilibrium thermodynamics[9], [10] and serves as the foundation for multiscale simulations and design of materials[5], [11], [12].

CALPHAD modeling of thermodynamics of multicomponent materials starts from the modeling of Gibbs energy of individual phases in their stable, metastable, and unstable states, first for pure elements and then moving up to binary and ternary systems. These binary and ternary models are then combined and extrapolated to describe complex alloys such as steels or nickel-base superalloys. Since these databases are built up from the pure elements, accurate thermodynamic descriptions of pure elements are critical, and any of their changes would require subsequent remodeling of all related higher-order systems[13], [14]. Their importance prompted the CALPHAD community to create a standardized database of pure element thermodynamic descriptions published in 1991[15] , serving as the foundation of the second generation of CALPHAD modeling building on its first generation[8], [16] , and commonly referred as SGTE91 database. In this database the heat capacity of an element was described by polynomials that were designed to cover the available high temperature data above 298.15 K, as used in the NIST-JANAF table[17] but also informed by additional thermochemical measurements, and the Gibbs energy differences between stable and non-stable phases of pure elements, commonly referred as lattice stability[8], [16] , were defined.  This standardized database of pure elements has enabled the development of many valuable CALPHAD databases for thermodynamic, kinetic, and other properties of complex multicomponent materials that are widely used in academia and industry[11], [18] .

One of the major limitations of the second generation CALPHAD modeling is that it is applicable only to high temperatures and cannot be extended to low temperatures, where the heat capacity can no longer be described by simple polynomials, and physics-based models such as Debye or Einstein models are needed. On the other hand, tremendous amount of thermodynamic data has been predicted via first-principles calculations based on density functional theory (DFT)[19], [20] , and only part of them is usable in CALPHAD modeling[21]. Furthermore, since the SGTE91 database was developed more than 30 years ago, efforts have been made to improve CALPHAD modeling of pure elements, commonly referred to as the $3^{rd}$ generation CALPHAD models. These models started with the 1995 Ringberg workshop (RW)[22] where a "universal" model was proposed for CALPHAD modeling of pure elements down to 0 K using an Einstein function and several power terms.  Chen and Sundman (CS)[23] modified the RW model at high temperature and applied it to FCC, BCC, and liquid iron, and Roslyakova et al.[24]  proposed a segmented regression (SR) model to replace the polynomial function in the SGTE91 model.  There have been several publications on the development of $3^{rd}$ generation CALPHAD models with extensions to binary[25], [26], [27], [28], [29], [30], [31], [32], [33], [34], [35], [36]and ternary systems[37], [38]. In addition, three recent re-assessments employing the SR model have extended these approaches to complex high-order systems: the Mo–Nb–Ta–W refractory high-entropy alloys[39], Mo–Nb–Ta–W–Hf–Zr alloys[40], and Ti–Hf–Zr–Nb–Ta alloys[41].

These studies demonstrate that 3[rd] generation CALPHAD modeling is now actively used for re-assessments across a wide range of systems. However, the complete re-assessment of existing multicomponent databases would require many years of effort. This represents a major drawback for accelerating database development and for integrating more physics-based models into CALPHAD. It is noteworthy that, so far, only the SR model has been applied in re-assessments of multicomponent systems with more than four elements, underlining its importance for extending CALPHAD to high-order alloys.

Moreover, the lack of systematic revision of modeling of pure elements reflects an inherent feature of the hierarchical CALPHAD approach where the modeling of pure elements is at the foundation of binary, ternary and multicomponent systems, and any change of modeling of a pure element requires the re-modeling of all systems containing that element[13], [14] , which is practically intractable with existing data and tool infrastructure in the CALPHAD community. The key bottlenecks are the repository of input data and high throughput tools for CALPHAD modeling so various models could be evaluated and compared, and multicomponent CALPHAD databases could be efficiently updated[13], [14] . Recently, open-source tools PyCalphad[42], [43] and ESPEI[44], [45] have been developed with PyCalphad for thermodynamic models and calculations and ESPEI for efficient evaluations of model parameters using Markov Chain Monte Carlo (MCMC) simulations and JSON files for the repository of input data.  The default models of pure elements in PyCalphad and ESPEI are the SGTE91 database.

These open-source python tools are ideal for implementation of pure element models as they enable efficiency in evaluation of model parameters through automation. Additionally, as computational predictions from density functional theory (DFT) and machine learning (ML) methods improve, thermodynamic database development can potentially be automated from pure elements to multicomponent systems. DFT can predict Gibbs energy of configurations to enable the initial evaluations of model parameters, which are further optimized to reproduce experimental phase equilibrium data[21] . Though DFT predictions are often performed at 0 K, they can also be at finite temperature by including contributions from thermal electrons and phonons[46] . More recently, ML models have been rapidly developed with accuracy close to DFT-based calculations and much higher efficiency[47], [48], [49] . With the Helmholtz energy of both stable and metastable configurations predicted by DFT-based calculations, the multiscale entropy approach (recently termed zentropy theory) was developed to accurately obtain Gibbs energy of individual phases with emergent properties[50], [51], [52], [53] , enabling the autonomous regeneration of CALPHAD model parameters for individual phases with PyCalphad and ESPEI.

The present work implements the RW, CS, and SR models of pure elements into the PyCalphad/ESPEI ecosystem[54] and evaluates their model parameters for 41 pure elements using the available experimental heat capacity data in the literature. These experimental data are taken from the NIST Alloy Data repository[55].

## 2   Overview of Thermodynamic Models: SGTE91, RW, CS, and SR

As mentioned above, most CALPHAD software defaults to describing the temperature dependence of heat capacities with high-order polynomials fit to experimental data and then stored in the universally used SGTE91 database[15] for pure elements. The Gibbs energy function of a pure element in the nonmagnetic solid phase φ is described by the following equation:

$$^{0}G^{\varphi}(T) - H^{SER} = a + bT + cTln(T) + dT^2 + eT^{-1} + fT^3 + gT^7 + hT^{-9} \qquad Eq.\ 1$$

where $H^{SER}$ is the molar enthalpy of the element at 298.15 K and $10^5$ Pa in the standard element reference (SER) state so that $H = 0$ under the SER state.

The enthalpy, entropy, and heat capacity can be derived through the following relations:

$$S = -\left(\frac{\partial G}{\partial T}\right)_P = -(b + c(1 + \ln(T)) + 2dT - eT^{-2} + 3fT^2 + 7gT^6 - 9hT^{-10}) \qquad Eq.\ 2$$

$$H = G + TS = G - T\left(\frac{\partial G}{\partial T}\right)_P$$

$$= a - cT - dT^2 + 2eT^{-1} - 2fT^3 - 6gT^7 + 10hT^{-9} \qquad Eq.\ 3$$

$$C_P = \left(\frac{\partial H}{\partial T}\right)_P = -T\left(\frac{\partial^2 G}{\partial T^2}\right)_P = T\left(\frac{\partial S}{\partial T}\right)_P$$

$$= -c - 2dT - 2eT^{-2} - 6fT^2 - 42gT^6 - 90hT^{-10} \qquad Eq.\ 4$$

For magnetic elements, the heat capacity due to magnetic order-disorder transition is added separately as discussed below, and contributes to enthalpy, entropy, and Gibbs energy accordingly. To extend Eq. 1 down to 0 K, the 3rd generation models for heat capacity have been developed with three parts: the Einstein or Debye model for phonon contribution, a model for additional low and high temperature effects, and a magnetic model for magnetic elements in accordance with physical contributions discussed by Grimvall[56]. The RW model is written as follows[22],

$$C_p^{RW}(T, \theta^{RW}) = C_p^{Deb/Ein}(T) + aT + bT^2 + C_p^{Magn}(T) \qquad Eq.\ 5$$

where the Debye and Einstein models are given by following equations

$$C_p^{Deb}(T, \theta_D) = 9R\left(\frac{T}{\theta_D}\right)^3 \int_0^{\frac{\theta_D}{T}} \frac{x^4 e^x}{(e^x - 1)^2}\, dx \qquad Eq.\ 6$$

$$C_p^{Ein}(T, \theta_E) = 3R\left(\frac{\theta_E}{T}\right)^2 \frac{e^{\frac{\theta_E}{T}}}{\left(e^{\frac{\theta_E}{T}} - 1\right)^2} \qquad Eq.\ 7$$

where $\theta_D$ and $\theta_E$ are the Debye and Einstein temperature, respectively. The Debye model represents a better temperature dependence at low temperatures than the Einstein model, but the integration in the Debye model makes it difficult for its implementation in the current CALPHAD modeling tools though attempts have been made to represent it by a combination of two Einstein functions in several systems[25], [26], [27], [39], [40], [41].

The magnetic contribution to heat capacity has been described by the Inden-Hillert-Jarl model[57], which was subsequently revised by Chen and Sundman[23] and further refined by Xiong et al.[58]. Although the model by Xiong et al. was primarily developed for binary and multicomponent systems, its formulation is applicable to pure elements and is used as such in the present work. The present work adopts the formulation by Xiong et al. as follows:

$$C_p^{Magn} = RT * g(\tau) * \ln(\beta^* + 1) \qquad Eq.\ 8$$

$$g(\tau) = \begin{cases} \dfrac{0.63570895}{A}\left(\dfrac{1}{p} - 1\right)\left(2\tau^3 + 2\dfrac{\tau^9}{3} + 2\dfrac{\tau^{15}}{5} + 2\dfrac{\tau^{21}}{7}\right), \tau \le 1 \\ \dfrac{1}{A}\left(2\tau^{-7} + 2\dfrac{\tau^{-21}}{3} + 2\dfrac{\tau^{-35}}{5} + 2\dfrac{\tau^{-49}}{7}\right), \tau \ge 1 \end{cases} \qquad Eq.\ 9$$

$$A = 0.33471979 + 0.49649686\left(\frac{1}{p} - 1\right) \qquad Eq.\ 10$$

where $\tau$ equals T/T$^*$, with T$^*$ being the magnetic Curie or Néel temperature, $\beta^*$ is the effective magnetic momentum per atom, and $p$ is the structure factor defined as the ratio of the magnetic enthalpy in the paramagnetic state to the total magnetic enthalpy.

In the CS model developed for iron[23], the $T^2$ term in the RW model with Einstein model was replaced by $T^4$ as they found better agreement with high-temperature heat capacity data as follows:

$$C_p^{CS} = C_p^{Ein} + aT + bT^4 + C_p^{Magn} \qquad Eq.\ 11$$

In the SR model[24] segments are smoothly jointed in a temperature range by quadratic bends, as follows:

$$C_p^{SR} = C_p^{Ein} + C_p^{BCM} + C_p^{Magn} \qquad Eq.\ 12$$

$$C_p^{BCM} = \beta_1 T + \beta_2 * q(\tau,\gamma) \qquad Eq.\ 13$$

$$q(\tau,\gamma) = f(x) = \begin{cases} 0, & T < \tau - \gamma \\ \dfrac{(T-\tau+\gamma)^2}{4\gamma}, & \tau - \gamma \leq T \leq \tau + \gamma \\ T - \gamma, & T > \tau + \gamma \end{cases} \qquad Eq.\ 14$$

where the temperature range is defined by $\tau$-$\gamma$ and $\tau$+$\gamma$ with $\beta_1$ and $\beta_1 + \beta_2$ being the slopes at $\tau$-$\gamma$ and $\tau$+$\gamma$, respectively. It is noted that in the SR model, there are four fitting parameters in addition to those in $C_p^{Deb/Ein}$ and $C_p^{Magn}$, i.e., $\tau$, $\gamma$, $\beta_1$, and $\beta_2$, while in the RW and CS models, there are only two additional parameters, i.e., $a$ and $b$.

With the heat capacity model parameters evaluated, entropy, enthalpy, and Gibbs energy can be obtained by integrations of Eqs. 15-17 as follows:

$$S = \int_0^T \frac{C_p}{T} dT \qquad\qquad Eq.\ 15$$

$$H = \int_0^T C_p dT - H_0 \qquad\qquad Eq.\ 16$$

$$G = H - TS \qquad\qquad Eq.\ 17$$

where $H_0$ represents the enthalpy at 298.15 K so that $H$=0 at the SER state used in CALPHAD modeling. The present work implemented the RW, CS, and SR models into PyCalphad using the Einstein model to compare their results. The model parameters were evaluated using ESPEI with the same data and the same procedure. Other models can be added and compared accordingly when available.

The goal of the present work is not to propose new thermodynamic models, but to systematically implement, compare, and quantify uncertainty for existing pure element models within a unified PyCalphad and ESPEI framework. By applying identical data, fitting procedures, and statistical metrics to the RW, CS, and SR models across 41 elements, this work enables transparent model comparison and provides a scalable foundation for automated CALPHAD database regeneration and future integration of physics based models

## 3   Brief Overview of PyCalphad and ESPEI

As previously mentioned, revising a pure element within a multicomponent system requires extensive re-modeling of many systems, as illustrated in Figure 1. Automating this reassessment process would significantly facilitate the development of new models. PyCalphad and ESPEI provide a foundation for this automation.

*Figure 1 Circled systems to be remodeled in a hypothetical 6 element database of A-B-C-D-E-F, if the thermodynamic description of pure Element "C" is modified, reproduced from Ref.* [13].

PyCalphad is an open-source Python library for creating thermodynamic models, calculating thermodynamic properties and phase diagrams, and investigating driving forces and phase equilibria[42] . It plays a significant role in computational thermodynamics by providing a flexible and powerful framework for modeling complex systems. PyCalphad is designed with models represented symbolically by leveraging symbolic mathematics to define and manipulate thermodynamic models. This greatly facilitates the inclusion of additional custom models.

ESPEI is a Python tool built on PyCalphad for evaluating model parameters and uncertainty of CALPHAD models[44]. ESPEI operates in two stages: estimation and optimization of model parameter values. During the estimation stage, ESPEI employs a linear fitting strategy to parameterize Gibbs energy functions of single phases based on their thermochemical data, which are derivatives of Gibbs energy. In the optimization stage, ESPEI refines the model parameters using phase equilibrium data between phases through Bayesian parameter evaluation with MCMC simulations. This two-stage process ensures that the models are both accurate and robust, providing reliable predictions for thermodynamic properties.

In the estimation stage, ESPEI uses a corrected Akaike information criterion (AICc)[59] to compare modeling results for a given dataset with penalties for the number of parameters to prevent overfitting. The function for AIC is given as:

$$AIC = -2 \ln L + 2k,$$
*Eq. 18*

where L is the likelihood and k the number of parameters. The log likelihood under maximum likelihood estimation (MLE) with normally distributed errors is:

$$\ln L_{MLE} = -\frac{n}{2} \ln \frac{RSS}{n} + C.$$
*Eq. 19*

Here, RSS is the sum of square residuals between observed and predicted data, and C is a constant[60]. Since C is constant for all models on a dataset, it can be ignored when just comparing one dataset. As such AIC becomes:

$$AIC = n \ln \frac{RSS}{n} + 2k$$
*Eq. 20*

The AICc used in ESPEI adds further penalization to models with higher numbers of parameters especially in the presence of small amounts of data[59]. This AICc is as follows:

$$AICc = AIC + \frac{2k^2 + 2k}{n-k-1} = n \ln \frac{RSS}{n} + 2k + \frac{2k^2 + 2k}{n-k-1} = n \ln \frac{RSS}{n} + \frac{2kn}{n-k-1}$$
*Eq. 21*

where n represents the number of data points.

To perform Bayesian optimization, ESPEI uses an ensemble MCMC sampler to explore the parameter space. Each individual Markov chain independently samples from the posterior distribution of model parameters. Each chain evolves according to the MCMC algorithm, generating a sequence of parameter sets that reflect the likelihood of those values given the data and prior assumptions. The use of multiple walkers enhances the diversity of sampling paths, helps ensure more comprehensive exploration of the parameter space, and reduces the likelihood of chains becoming trapped in local minimum. As the chains evolve, they generate a distribution of parameter values that reflect the uncertainty informed by both the prior assumptions and the experimental data. These distributions form the basis for uncertainty quantification (UQ) and propagation in the final thermodynamic models.

Uncertainty of the input data and from the MCMC chains are used for UQ of parameters. Uncertainty in input data can be directly included when available, and ESPEI uses a default value of 0.2 J/K mol for heat capacity data when uncertainty is not available. The uncertainty in parameter values is represented by their distribution across converged multiple chains. Convergence can be assessed by means of several criteria common to MCMC simulations, as detailed by Gelman et al.[61]. The uncertainty propagation, particularly to higher order systems, was investigated by Paulson et al. [62], who further considered two weighting methods to quantify uncertainty of experimental data sets used in the $Cp$-data of Al and Hf with the SR model[63].

## 4   Modeling Procedure

The SGTE91 database contains 81 elements. 41 pure elements were considered in the present work with data collected so far, and extension to all elements in the periodic table is being considered. Of the 41 elements, the majority were directly collected from the alloy database maintained by the Thermodynamics Research Center (TRC) at the National Institute of Standards and Technology (NIST)[55]. It is curated by experts with a focus on the elements, binary and ternary alloys from open literature spanning from the early 1900s to recent publications, including properties such as density, specific heat, viscosity, enthalpy, phase transition temperatures, and more. Regardless of whether the original data are presented in tables, equations, or graphs, they are digitized and accompanied by metadata, provenance details, and uncertainty assessments. The database is accessible via a user-friendly web application[55] that offers both a graphical periodic table interface for simple searches and advanced options for complex queries with additional data analysis tools. Programmatic access is also available through an API, enabling efficient use of the structured data in electronic formats.

In the present work, pure element heat capacity data was systematically collected from the NIST database and converted to JSON format for maximum compatibility with ESPEI. Additionally, data for Al[24], Cr[64], and Fe[24] collected in previous publications and for Ge, Se, and Si curated in the present work were also utilized due to comprehensive analysis from various sources and compared with those from NIST database. All data used in the work is available in the GitHub repository[65] associated with this work. Heat capacity data in JSON format is the only mandatory input for fitting, as all models as well as the functionality are already implemented.

Models were manually added to PyCalphad, using symbolic representation of Eq. 5, Eq. 11, and Eq. 12 and their respective Gibbs energy functions. Magnetic contributions, where applicable, are incorporated using parameters such as Curie temperature, $\beta^*$, and p, which are automatically retrieved from PyCalphad and can be adjusted if necessary.

Bayesian parameter estimation is carried out using Markov Chain Monte Carlo (MCMC) sampling applied to the symbolic model functions. Multiple chains are employed to ensure thorough sampling and reduce the risk of local minima. In this work, 48 chains were run for 10,000 steps, with the initial 1,000 steps discarded as burn-in. Convergence is assessed before extracting parameter distributions, and uncertainties are quantified from the variance of converged chains. Median values of the posterior distributions are used as best-fit parameters. This can be seen in Figure 2, with the results of running the MCMC process on Cr with the CS model, with both the chains convergence and resulting corner plots results. A corner plot displays the marginal distributions of individual parameters as one-dimensional histograms along the diagonal, while the off-diagonal panels show two-dimensional density plots illustrating correlations between parameter pairs. This visualization helps assess parameter uncertainty, convergence, and interdependence in MCMC sampling results. From this corner plot, we can see that parameters $a$ and $b$ exhibit a moderate negative correlation, while $\theta_E$ shows weaker correlation with the other parameters, and all distributions appear unimodal and well-converged.

*Figure 2 Convergence of MCMC chains for parameters $\theta_E$, $a$, and $b$(top) and the corresponding corner plot (bottom) showing posterior distributions and parameter correlations for Cr using the CS model.*

The optimized models can also be validated by comparing calculated thermodynamic properties, such as Gibbs energy, enthalpy, and entropy, with established databases. Model performance is further evaluated using AICc, computed from the log-likelihood of the data given the fitted parameters. Results, including best-fit parameters and performance metrics, are stored in structured output files for subsequent analysis and comparison across elements.

This workflow provides a robust framework for integrating curated experimental data with advanced CALPHAD modeling techniques, enabling systematic parameter optimization and uncertainty quantification for pure elements.

## 5    Modeling Results and Discussions

Among the 41 elements studied, detailed results for four representative elements (Al, Cr, Zn, Fe) are presented in the main text, while results for the remaining elements are provided in the supplementary materials. They are selected to showcase face centered cubic (FCC, Al), body centered cubic and magnetic (BCC, Cr and Fe), and hexagonal close packed (HCP, Zn) elements, respectively. The present work focuses on modeling configurations stable at 0 K, i.e., the ground-state configuration of each element. The modeling of phases stable at high temperatures and liquid will be modeled in our future work. Einstein temperatures ($\theta_E$) of the 41 elements from the three models are listed in Table 1. Their AICc are listed in Table 2. It should be emphasized that the fitted Einstein temperature obtained in the present work is an effective parameter optimized to reproduce experimental heat capacity data over a finite temperature range, rather than a direct physical Debye temperature derived from elastic constants or phonon dispersion measurements. As a result, deviations from values estimated from Debye temperatures reported in the literature are expected, particularly for elements with limited low temperature data, strong elastic anisotropy, or significant anharmonic effects. Consequently, agreement with tabulated reference values should not be interpreted as a measure of fitting quality, and discrepancies do not indicate deficiencies in the modeling procedure. All input data and scripts are provided in the supplementary materials and are also available in Jupyter notebooks within the GitHub documentation[65].

*Table 1 Einstein parameter θE in kelvin evaluated for three models and compared to θE recalculated from θE ≅ 0.77θD [56] collected from Kittel [66]*

## 5.1 FCC Elements

The ground-state configuration of 12 pure elements is FCC, including Ag, Al, Au, Ca, Cs, Cu, Ir, Ni, Pb, Pd, Pt, Rh. Al was used as an example case for FCC, chosen for its common use and representative behavior of FCC elements. The only major outlier in FCC is the segmented regression model for pure Ca. Although heavy manual alterations could lead to a successful fit, the success of two other models indicated that the data was of sufficient quality for modeling, so further adjustments were not pursued.

The low accuracy at very low temperatures, where all models underestimate the heat capacity compared to experimental data, stems from the Einstein model's assumption that all atoms vibrate at a single frequency, which oversimplifies the true phonon spectrum of a solid. As Kittel describes, this leads to a failure to reproduce the characteristic $T^3$ dependence of heat capacity at low temperatures[66] . While the Debye model addresses this by incorporating a full distribution of vibrational modes, its implementation is more complex due to the required numerical integration.

Overall, the models show good agreement with experimental data and the SGTE91 Gibbs energies as seen in Figure 3, where all models shown are indistinguishable from each other for heat capacity, enthalpy, entropy, and Gibbs energy. The experimental data is shown in the heat capacity plot, where darker points indicate overlapping values. The spread in the data highlights experimental variability and underscores the importance of incorporating uncertainty in model predictions. It can be noted that as the temperature increases the Gibbs energy diverges from the SGTE91 Gibbs model. The underestimation of heat capacity at low temperatures is also evident, particularly below 100 K, where the models noticeably underpredict experimental values as discussed above.

*Figure 3 Thermodynamic properties of Al with three models: (a) Heat capacity with experimental data superimposed; (b) ΔEnthalpy, (c), Entropy, (d) Gibbs energy, and (e) Gibbs energy difference with respect to the SGTE91 database.*

## 5.2   BCC Elements

10 elements with the BCC ground-state configuration are modeled in the present work, i.e., Cr, Fe, K, Mo, Na, Nb, Rb, Ta, V and W. Cr was chosen as an example for BCC with heat capacity, enthalpy, entropy and Gibbs energy shown in Figure 4. Figure 4 highlights an important trend that occurs frequently in the heat capacity of the CS model. As shown in Figure 4a, the heat capacity starts to increase rapidly around the melting point due to the $T^4$ term. This trend is also present in the RW and SR models, but to a lesser degree. The modeling of heat capacity of solid phases above melting temperature is actively researched in the community[64], [67] and is being dealt with in our ongoing work. In addition, the antiferromagnetic behavior of Cr, which is known to influence thermodynamic properties at low temperatures, is not considered in the present assessment.

The three models differ noticeably in their descriptions, with a higher variation than what is seen in FCC Al. This increased variance is primarily due to the scatter in high-temperature experimental data, which prevents a clear trend and complicates the fitting process. Since the resulting fits of the three models are no longer nearly identical and more significant differences arise among them, a quantitative model comparison is useful for determining which model best suits the element.

*Figure 4 Thermodynamic properties of Cr with three models: (a) Heat capacity with experimental data superimposed; (b) ΔEnthalpy, (c) Entropy, (d) Gibbs energy, and (e) Gibbs energy difference with respect to the SGTE91 database*

## 5.3    HCP Elements

11 elements with the HCP ground-state structure were modeled in the present work, including. Be, Cd, Co, Hf, La, Mg, Re, Sc, Tl, Y, and Zn. Zn was selected to demonstrate as shown in Figure 5. All models behave similarly for Zn at lower temperatures but begin to diverge as temperature increases. This behavior is expected because differences in model formulations become more influential at elevated temperatures. Since Zn has a relatively low melting point and correspondingly low Debye temperature, the vibrational contribution remains significant across a broader portion of the temperature range compared to elements with higher Debye temperatures. This may explain a notable feature in Figure 5, where after the underprediction by the Einstein model between 5 and 10 K, Zn exhibits a visually apparent overprediction between 50 and 100 K. This occurs because the Einstein model's contribution increasing rapidly whereas the experimental values rise more gradually with temperature.

Overall, model performance for HCP elements is more varied compared to the high consistency seen in FCC and BCC elements. This variability is largely due to their limited and often incomplete experimental heat capacity data. For example, experimental data for Be was only available for up to 300 K, so fitting was restricted to that range. This introduces high uncertainty in the model's predictions at higher temperatures. Despite these challenges, the models can still perform well for HCP elements when the available data, even if limited, is internally consistent and spans a temperature range that is physically meaningful, capturing the dominant vibrational behavior at low temperatures and extending sufficiently toward higher temperatures to constrain the model parameters effectively.

*Figure 5 Thermodynamic properties of Zn with three models: (a) Heat capacity with experimental data superimposed; (b) ΔEnthalpy, (c) Entropy, (d) Gibbs energy, and (e) Gibbs energy difference with respect to the SGTE91 database*

### 5.4 Magnetic Elements

Four elements were modeled with magnetic contributions by Eq. 8, Eq. 9, and Eq. 10: HCP-Co, BCC-Fe, α-Mn, FCC-Ni. The results for Fe can be seen in Figure 6. The magnetic peak is clearly represented in the heat capacity, which is carried through the S and H curves. It reflects the loss of magnetic ordering near the Curie temperature. Since the magnetic contribution is independent of the parameter fitting, it adds the same contribution to all 3 models. As a result, the UQ present for these relects only uncertainties associated with the nonmagnetic $C_p$.

Since BCC-Fe transforms to FCC-Fe at 1185 K, and the current tool does not automatically handle solid-solid phase transformation, only experimental heat capacity data below this temperature is included in the diagram, showing good agreement between calculated and experimental data. The calculated $C_P$, H, S, and G agree with those from the SGTE91 description.

While the magnetic systems exhibited similar overall behavior, the differences between our results and SGTE values were consistently larger for magnetic materials than for nonmagnetic materials. However, despite exhibiting the greatest difference from SGTE, the magnetic contribution based on Xiong et al.[58] reproduces the experimental behavior across all four elements in all three models.

*Figure 6 Thermodynamic properties of Fe with three models: (a) Heat capacity with experimental data superimposed; (b) ΔEnthalpy, (c) Entropy, (d) Gibbs energy, and (e) Gibbs energy difference with respect to the SGTE91 database*

## 5.5    Other Elements

Seven additional elements were modeled in the present work, including As, Bi, Ge, Hg, Sb, Sn, Si, of which FCC/BCC/HCP are not their ground-state configurations. They perform similarly to other systems with the only outlier being difficulties with the segmented regression model for Si. Like with all other elements, all data and fittings can be found in the GitHub repository

## 5.6    Model Uncertainty and comparison

Models were compared using the AICc parameter for each element as shown in Table 2. Lower AICc values indicate a statistical better fit, and the model with the lowest AICc value for each element has been printed in bold in Table 2.

*Table* 2 AICc values of RW, CS and SR models

Among 41 elements studied, the SR, CS, and RW models were found to be the best fit for 17, 14, and 10 elements, respectively. However, the differences between models are small, and other criteria that are being implemented to expand the comparison between models[68] will be used to further rank the models. In addition to model selection, UQ of the parameters was performed using MCMC simulations. As described in the procedure section, uncertainty arises both from the input data and from the spread of parameter values across converged chains. The resulting posterior distributions provide insight into the confidence of each parameter estimate. While not shown in detail here, these distributions can be used to propagate uncertainty into thermodynamic predictions[62], offering a more robust basis for model comparison and application.

The uncertainty of the parameters can be visualized in Figure 7, where the Cp of Cr was fitted with the SR model. The darker line represents the median (50[th] percentile) of the posterior distributions while the shaded region spans the 16[th] to 84[th] percentile range, capturing the central 68% of the uncertainty. As shown, uncertainty increases with temperature, particularly beyond the melting point due to the lack of experimental Cp data for solid Cr above its melting temperature.

*Figure 7 Heat capacity of BCC Cr from the SR model fit by the MCMC simulation with the black line for the 50th percentile and the grey region for the 16th to 84th percentile of parameter distributions.*

## 6    Conclusion

The three existing heat capacity models of pure elements from 0 K to high temperature are implemented to the open-source PyCalphad and ESPEI systems, enabling systematical comparison of their performance using the same set of data and procedures. This new capability is tested for 41 pure elements in the present work, demonstrating its efficiency for CALPHAD modeling with uncertainty quantification. New models could be easily added and compared with existing models along with any new theoretical and experimental data. The present capability is being extended to include models for solid phases above melting temperature and liquid phases above and below melting temperature and will be ultimately incorporated into an autonomous CALPHAD modeling framework.

## 7    Code and Data Availability

Instructions for the PyCalphad and ESPEI downloads as well as a Jupyter notebook for application of codes can be found at https://github.com/amr8004/PureElementPRL. All data shown in the current work is also available in the PureElementPRL github repository[65]. The experimental data is available from NIST Metals Alloy User Interface at

https://trc.nist.gov/metals_data/. If GitHub codespaces are available,

https://github.com/amr8004/3rdGenerationToolCodespace can be run and will enable use of the tool through GitHub's UI on web.

## 8    Acknowledgements

The authors thank Brandon Bocklund and Richard Otis for developing the open-source tools and for their support in extending and building upon their work. This work was funded by grant contract No. N00014-21-1-2608 from the Office of Naval Research. Irina Roslyakova acknowledges the financial support from the Collaborative Research Center Superalloys Single Crystal (SFB TR-103 project T2) of the German Research Foundation (DFG). We honor the memory of our co-author, Abdulmonem Obaied, who passed away recently.

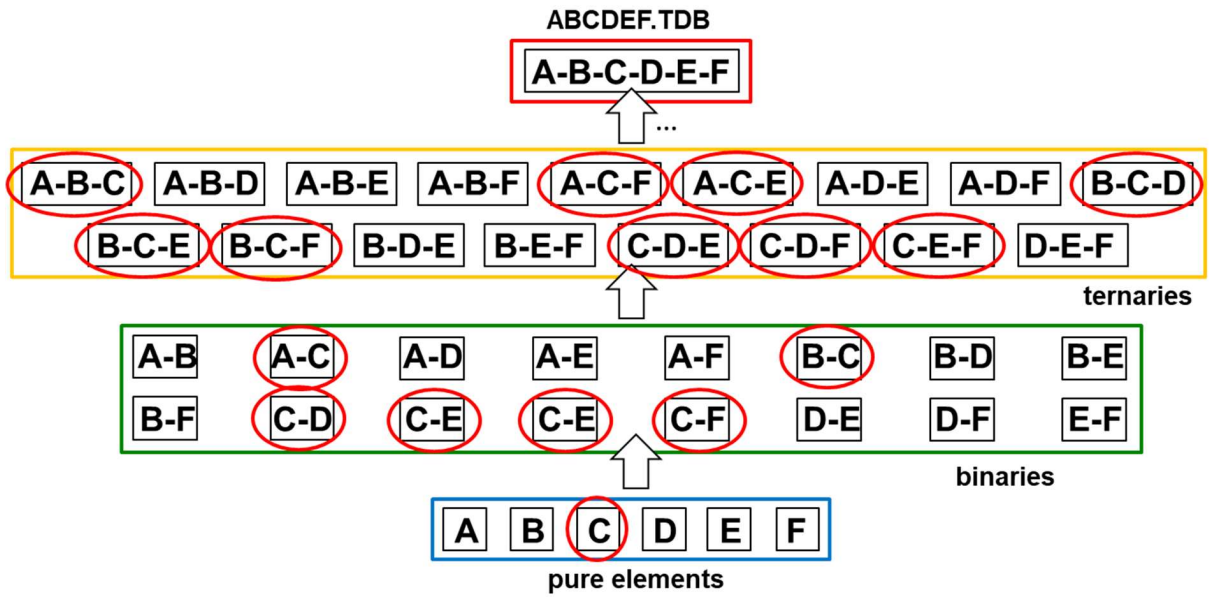

*Figure 1 Circled systems to be remodeled in a hypothetical 6 element database of A-B-C-D-E-F,*

*if the thermodynamic description of pure Element "C" is modified, reproduced from Ref.* [13].

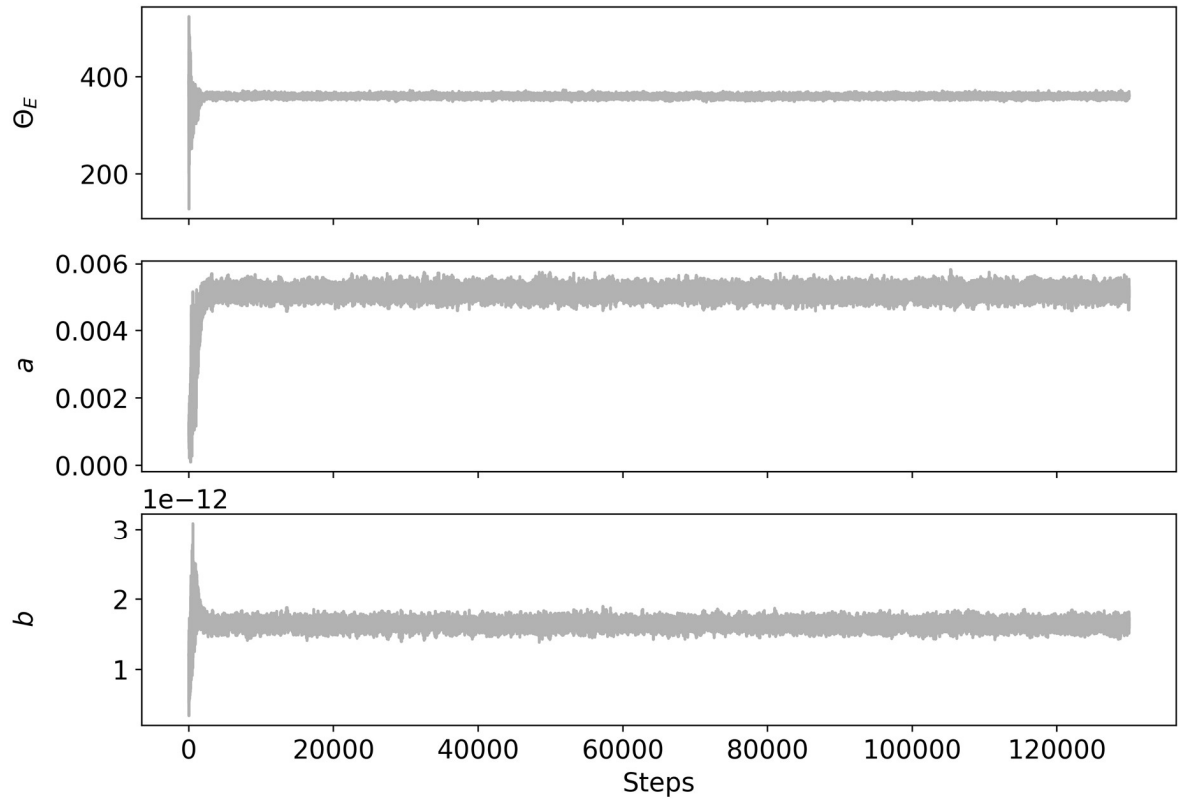

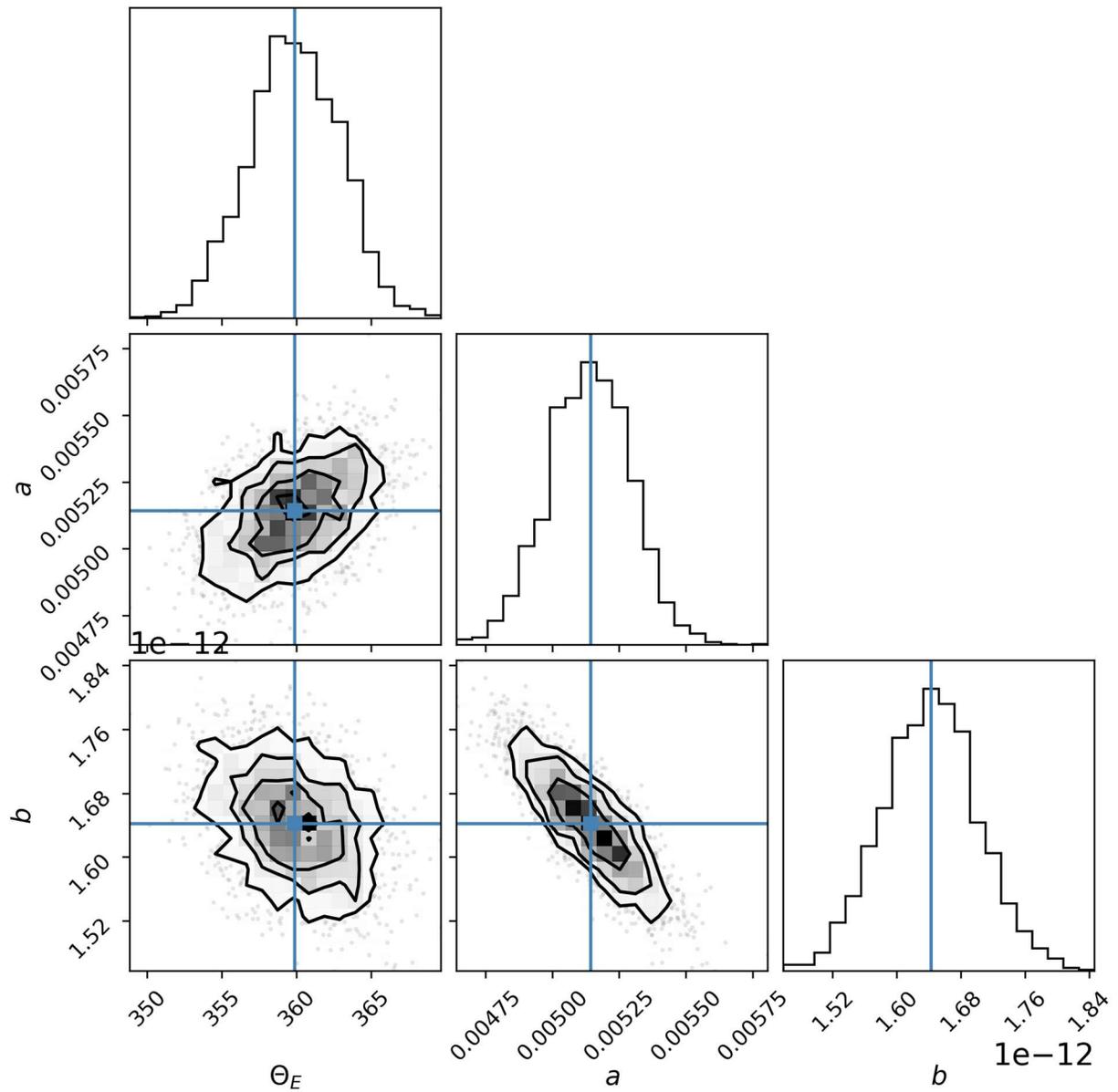

*Figure 2 Convergence of MCMC chains for parameters $\theta_E$, a, and b(top) and the corresponding corner plot (bottom) showing posterior distributions and parameter correlations for Cr using the CS model.*

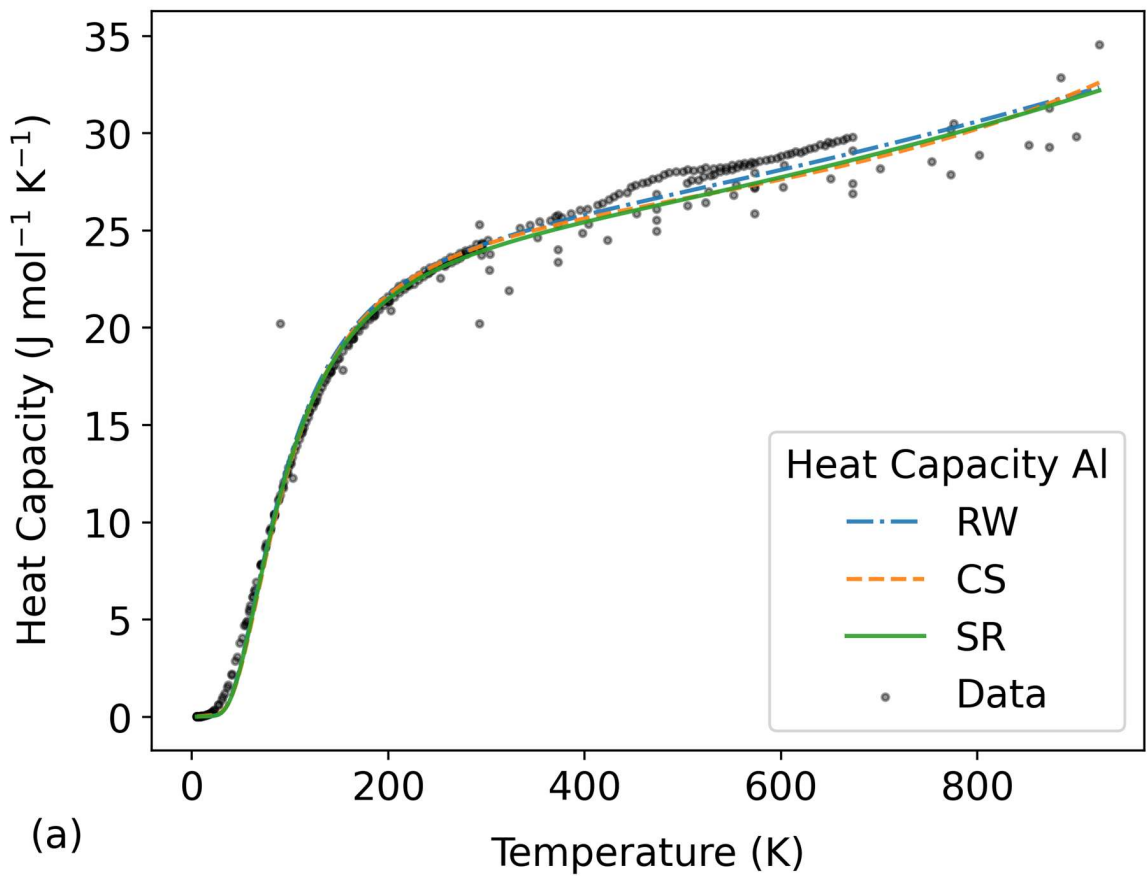

(a)

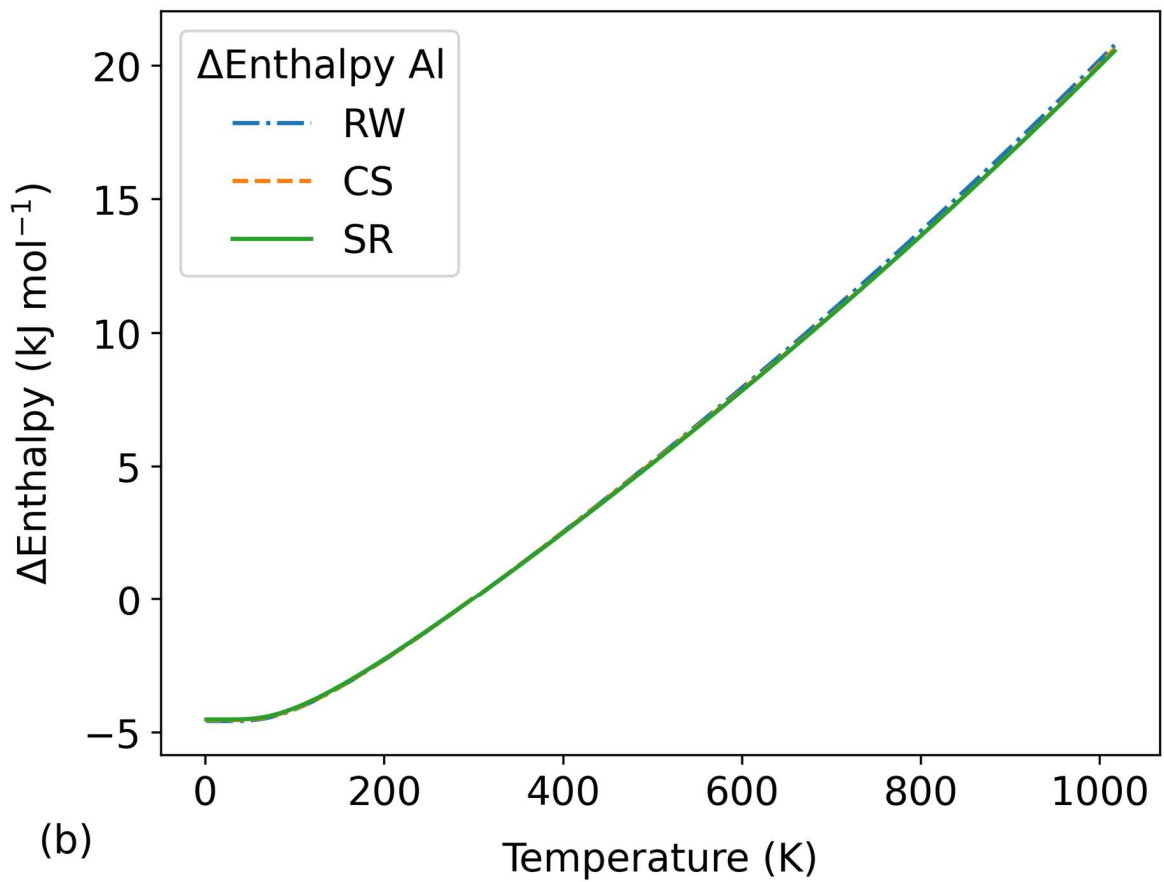

(b)

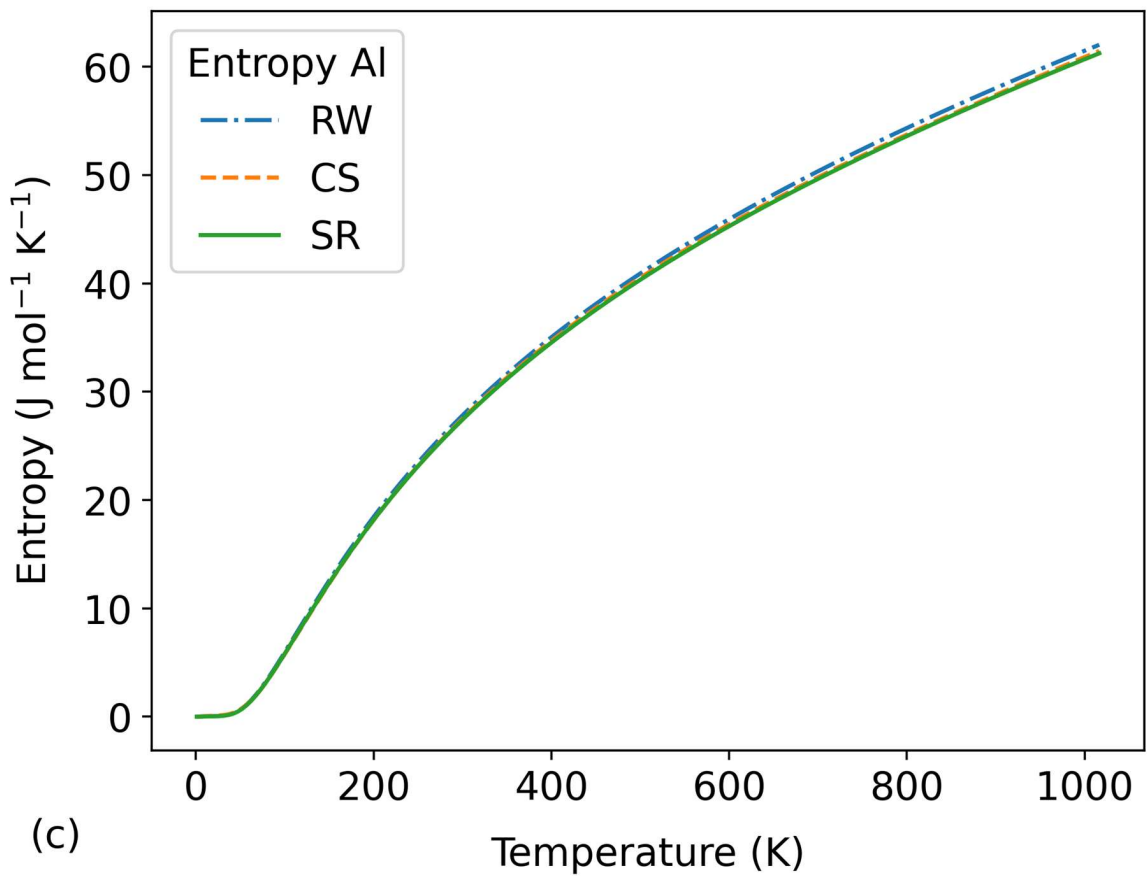

(c)

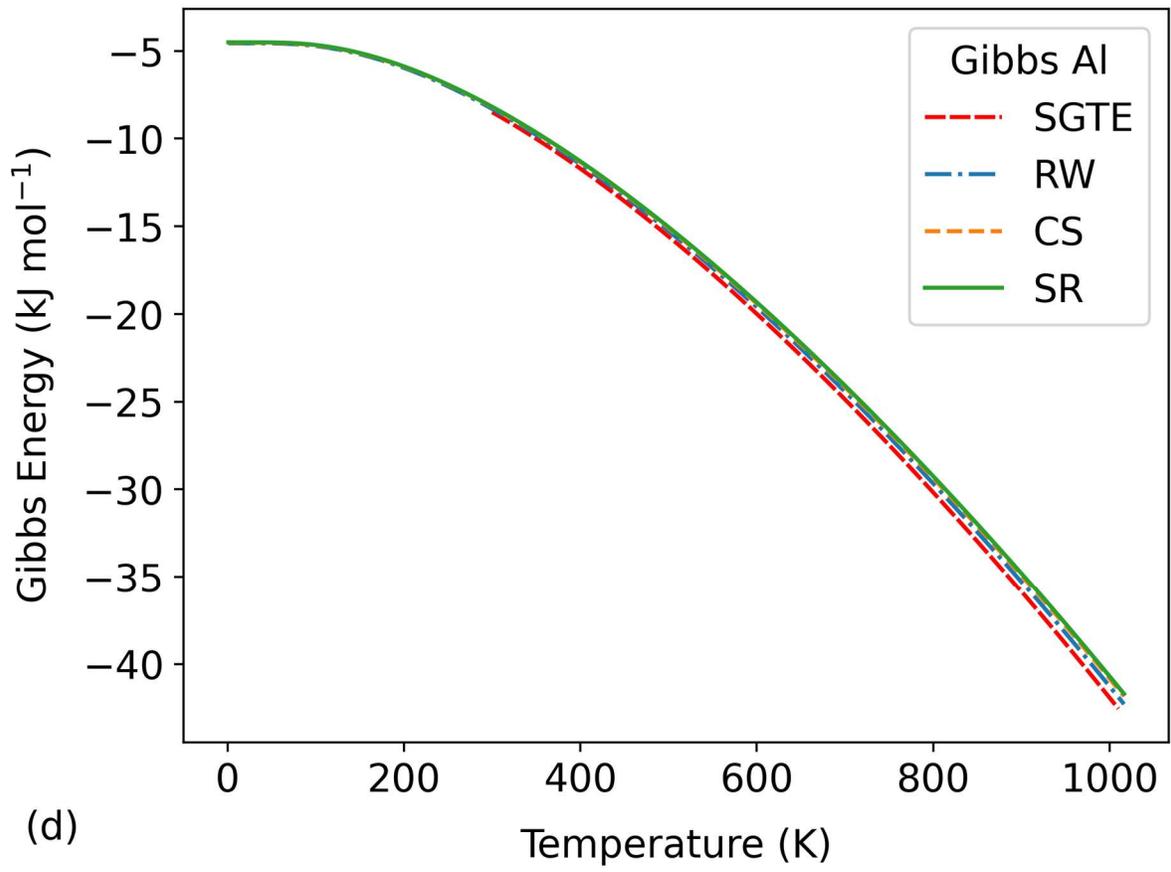

(d)

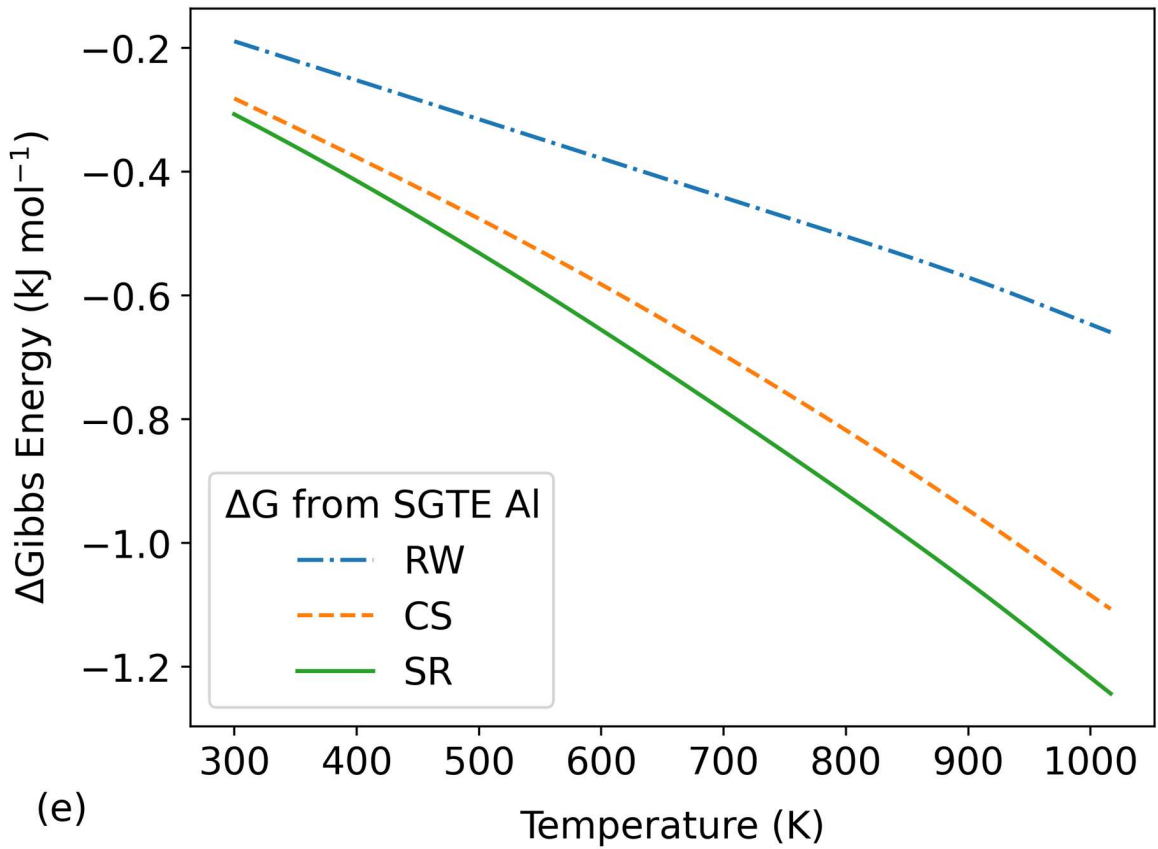

(e)

*Figure 3 Thermodynamic properties of Al with three models: (a) Heat capacity with experimental data superimposed; (b) ΔEnthalpy, (c), Entropy, (d) Gibbs energy, and (e) Gibbs energy difference with respect to the SGTE91 database.*

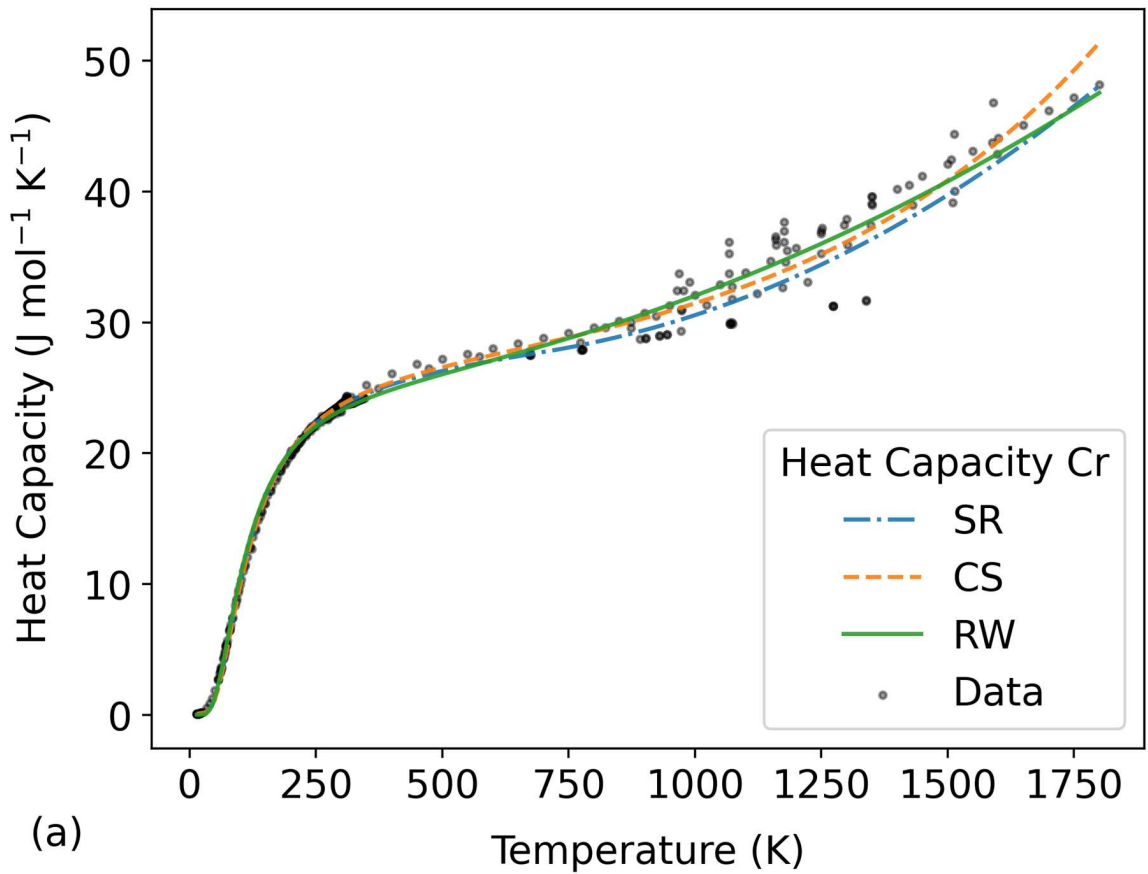

(a)

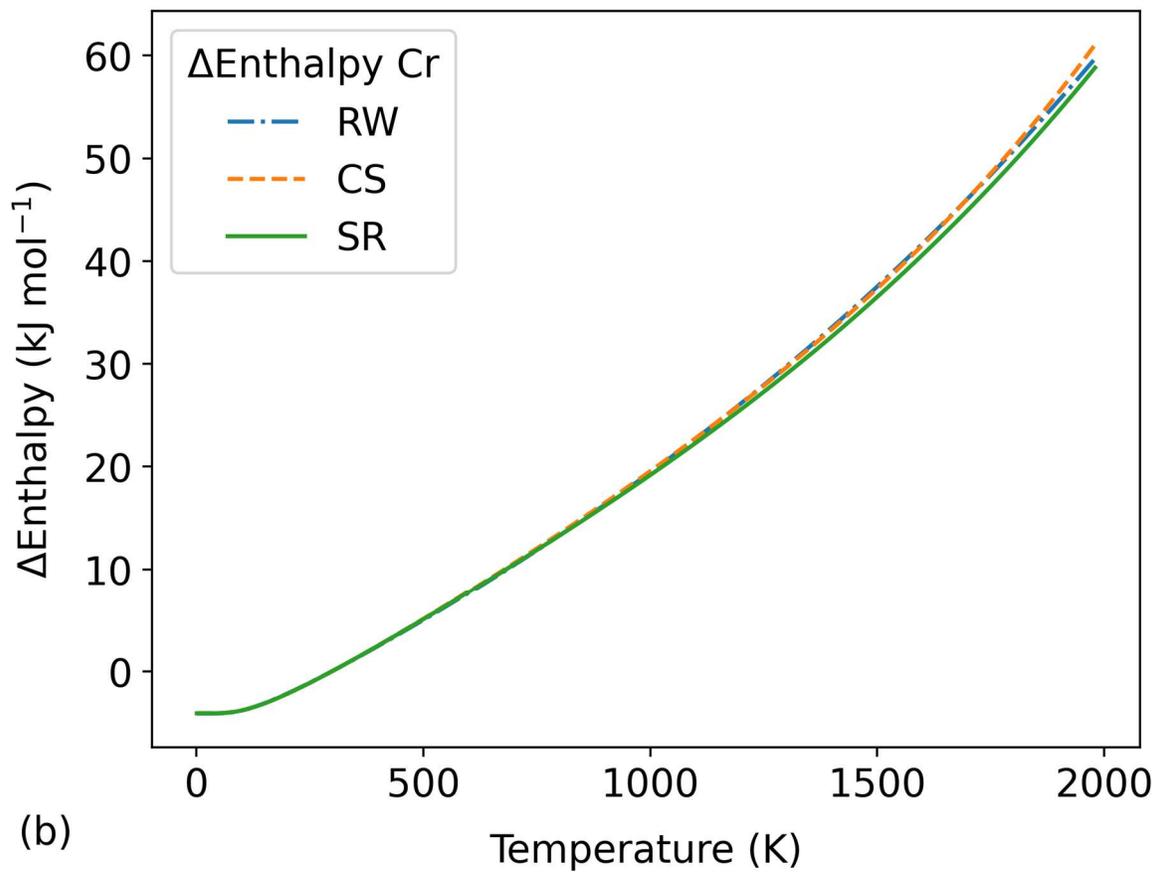

(b)

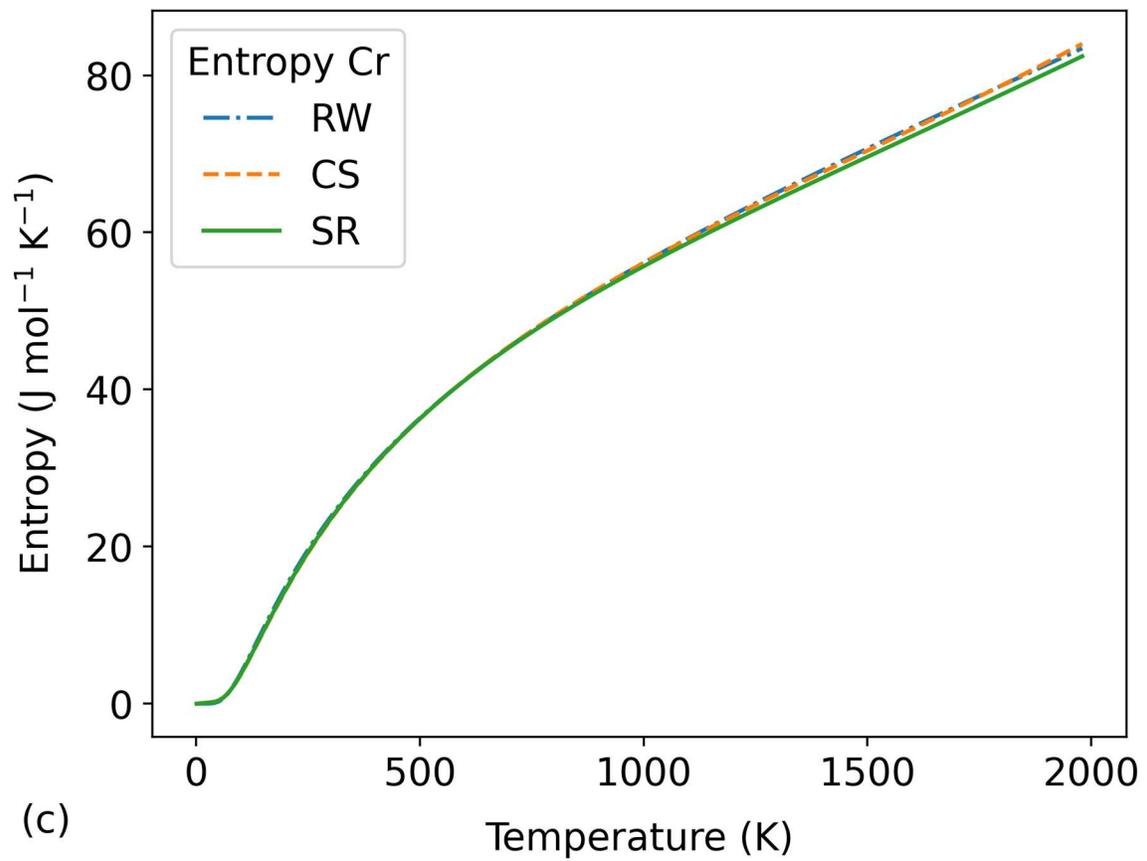

(c)

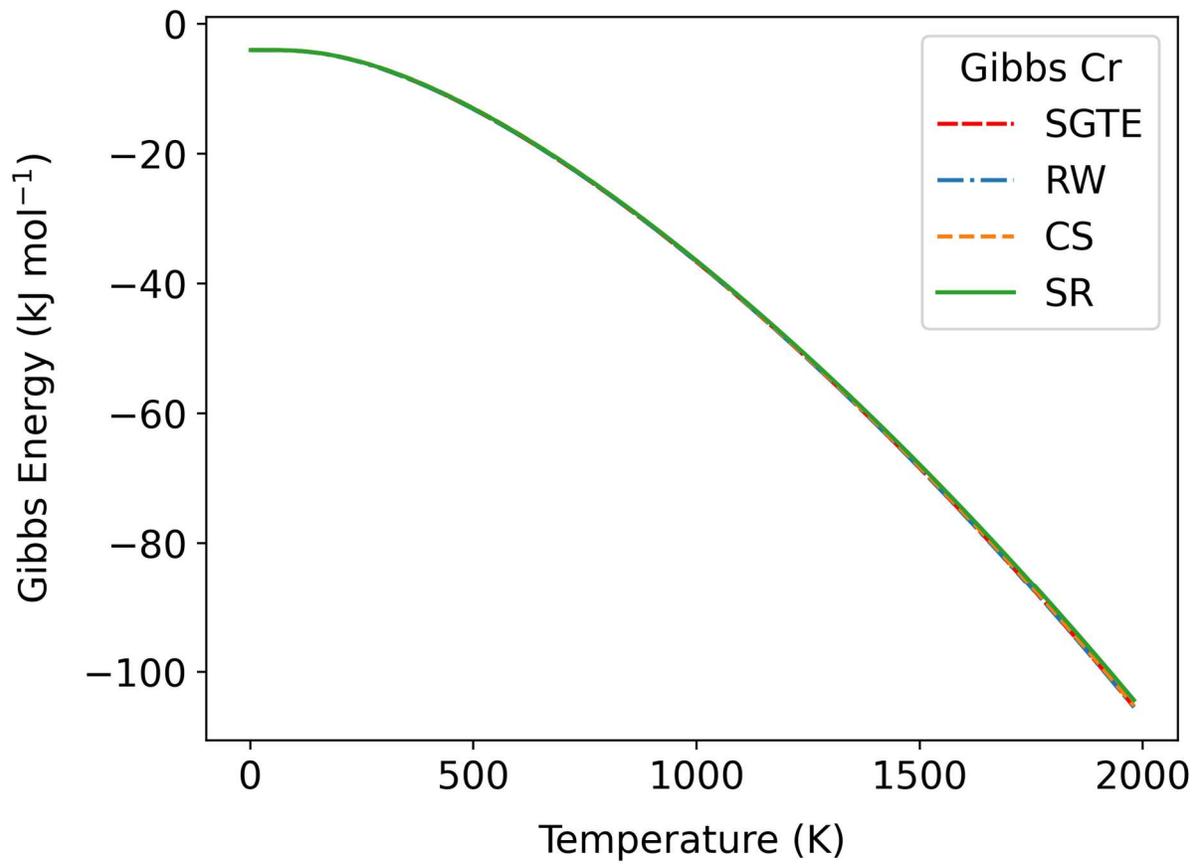

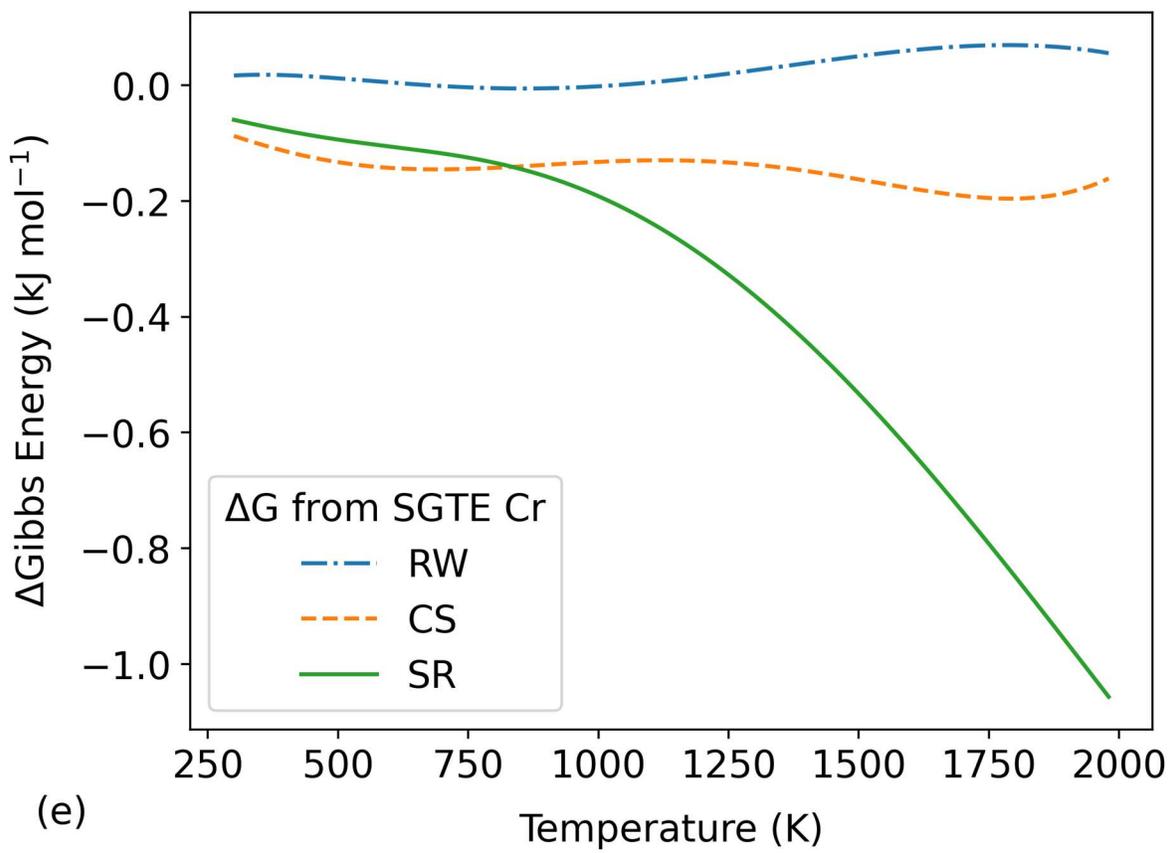

(e)

*Figure 4 Thermodynamic properties of Cr with three models: (a) Heat capacity with experimental data superimposed; (b) ΔEnthalpy, (c) Entropy, (d) Gibbs energy, and (e) Gibbs energy difference with respect to the SGTE91 database.*

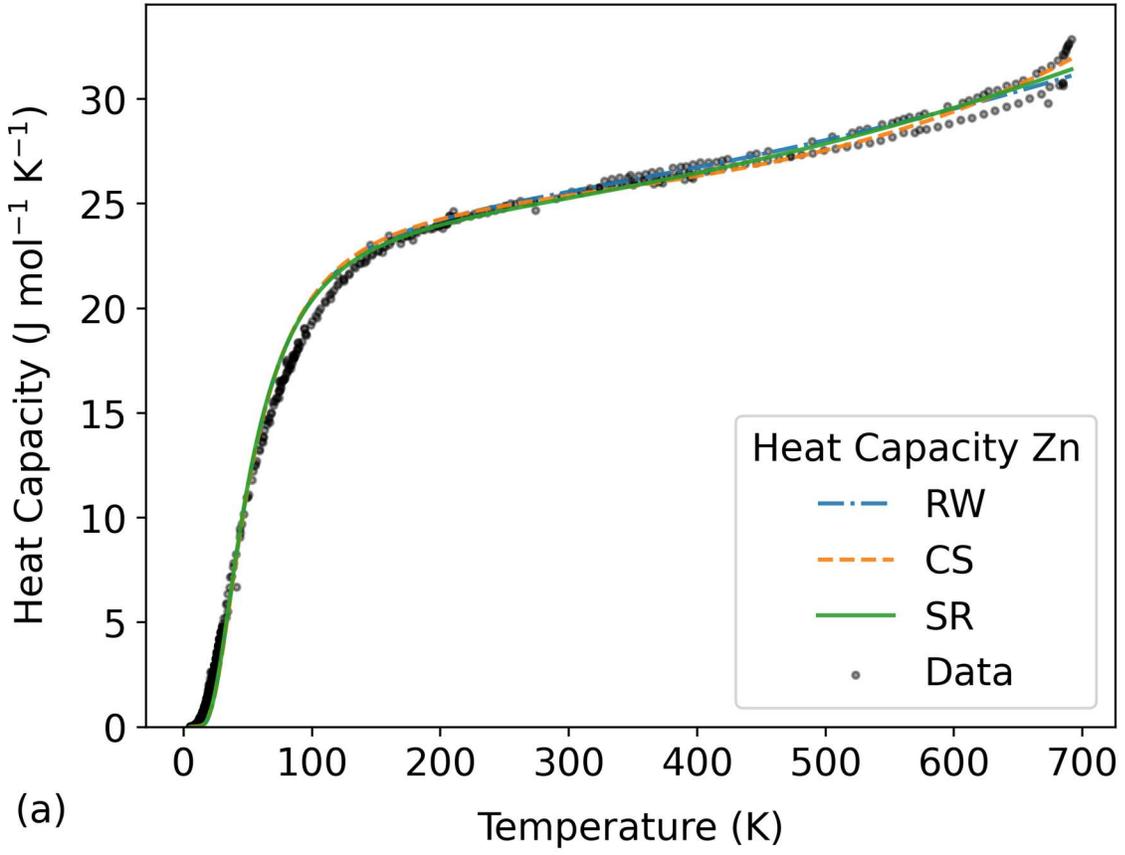

(a)

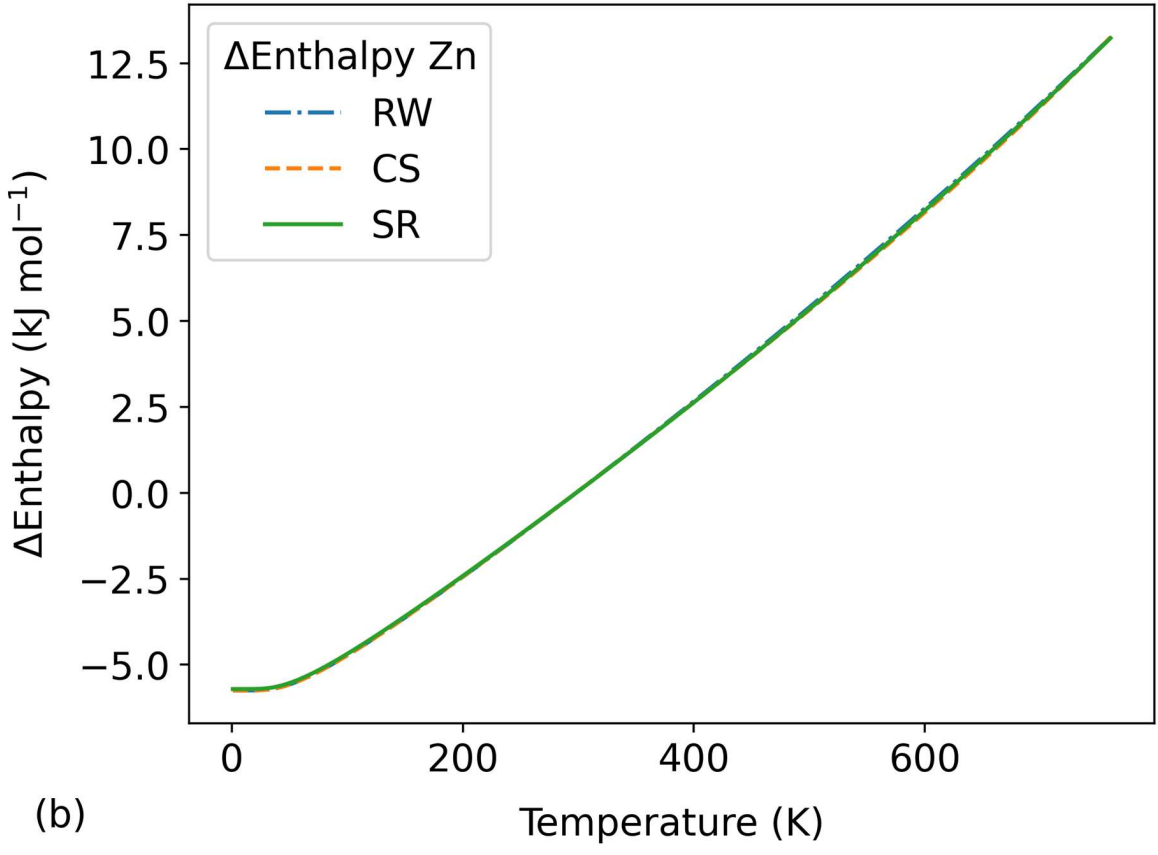

(b)

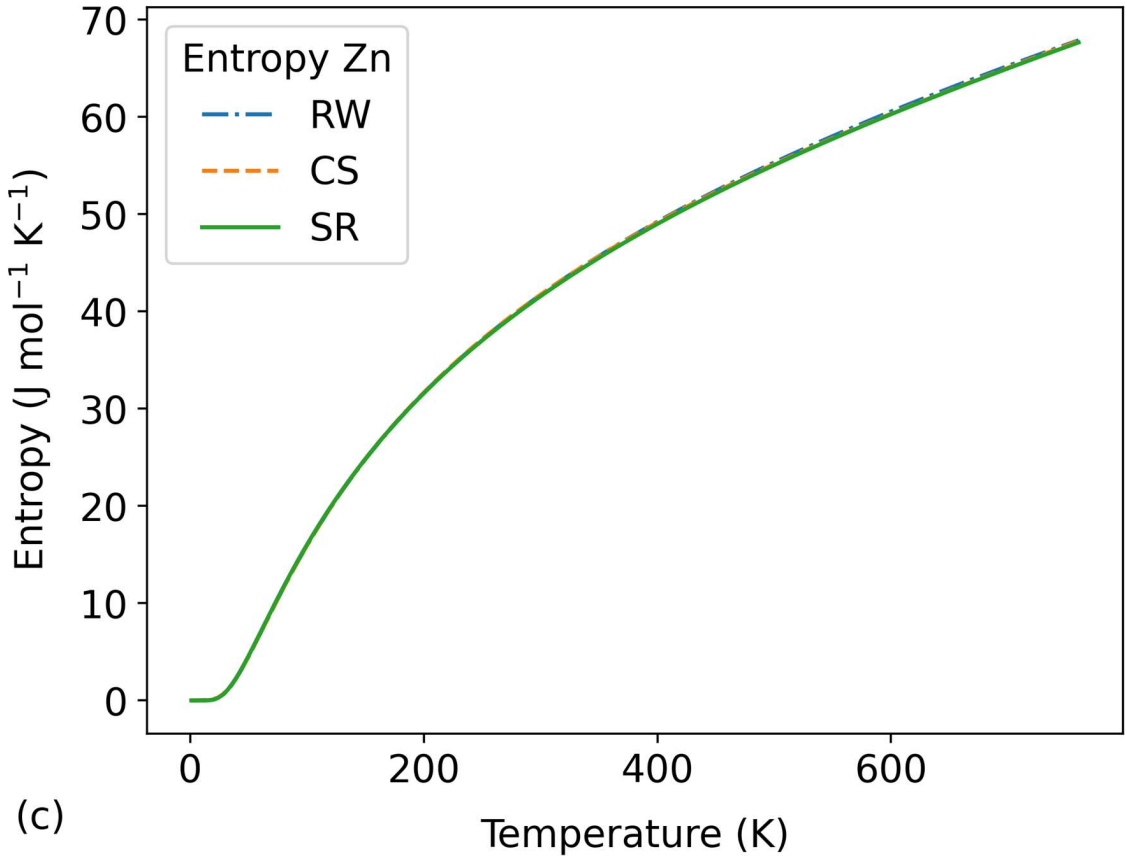

(c)

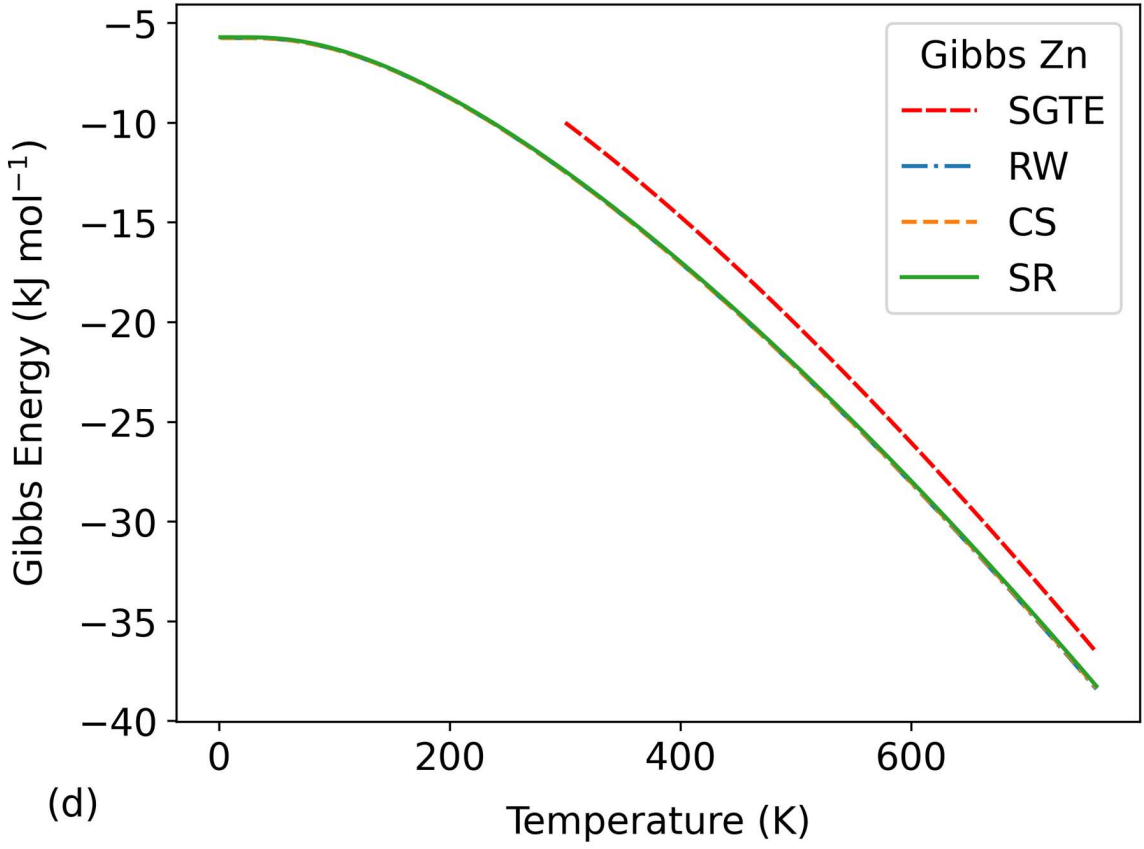

(d)

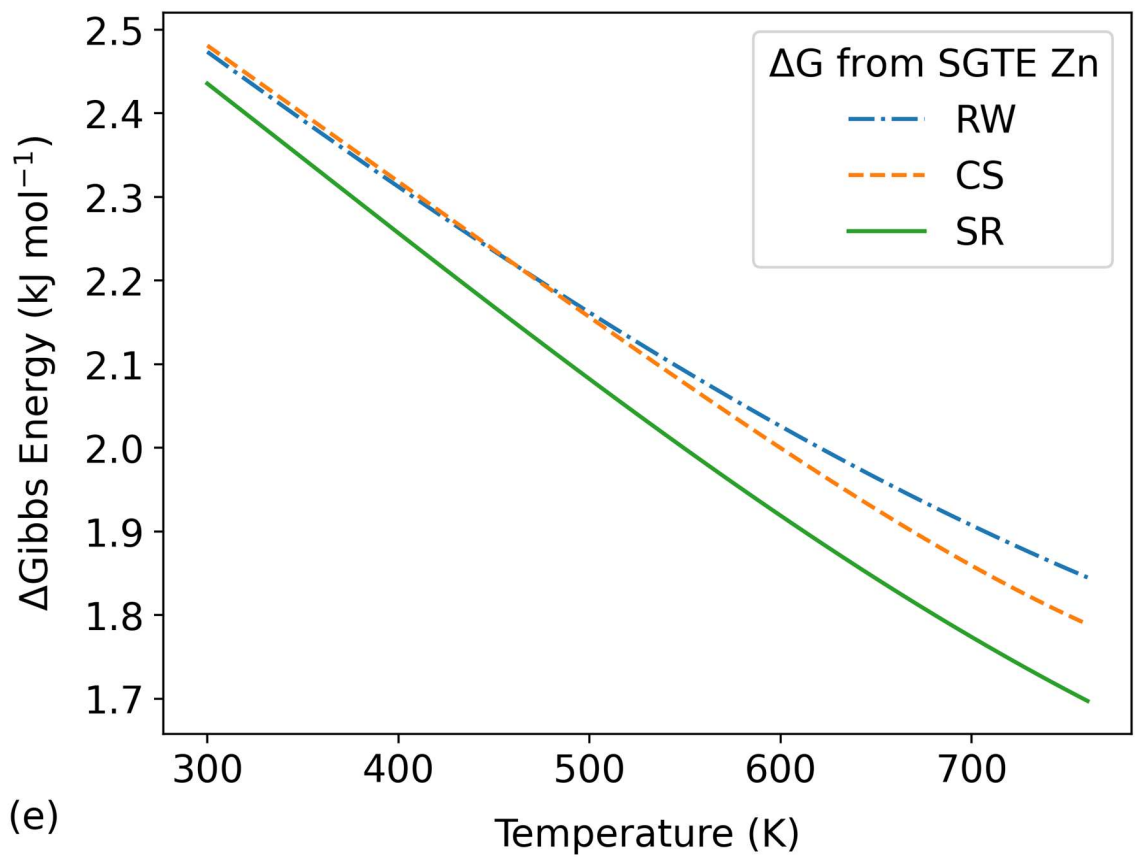

(e)

*Figure 5 Thermodynamic properties of Zn with three models: (a) Heat capacity with experimental data superimposed; (b) ΔEnthalpy, (c) Entropy, (d) Gibbs energy, and (e) Gibbs energy difference with respect to the SGTE91 database*

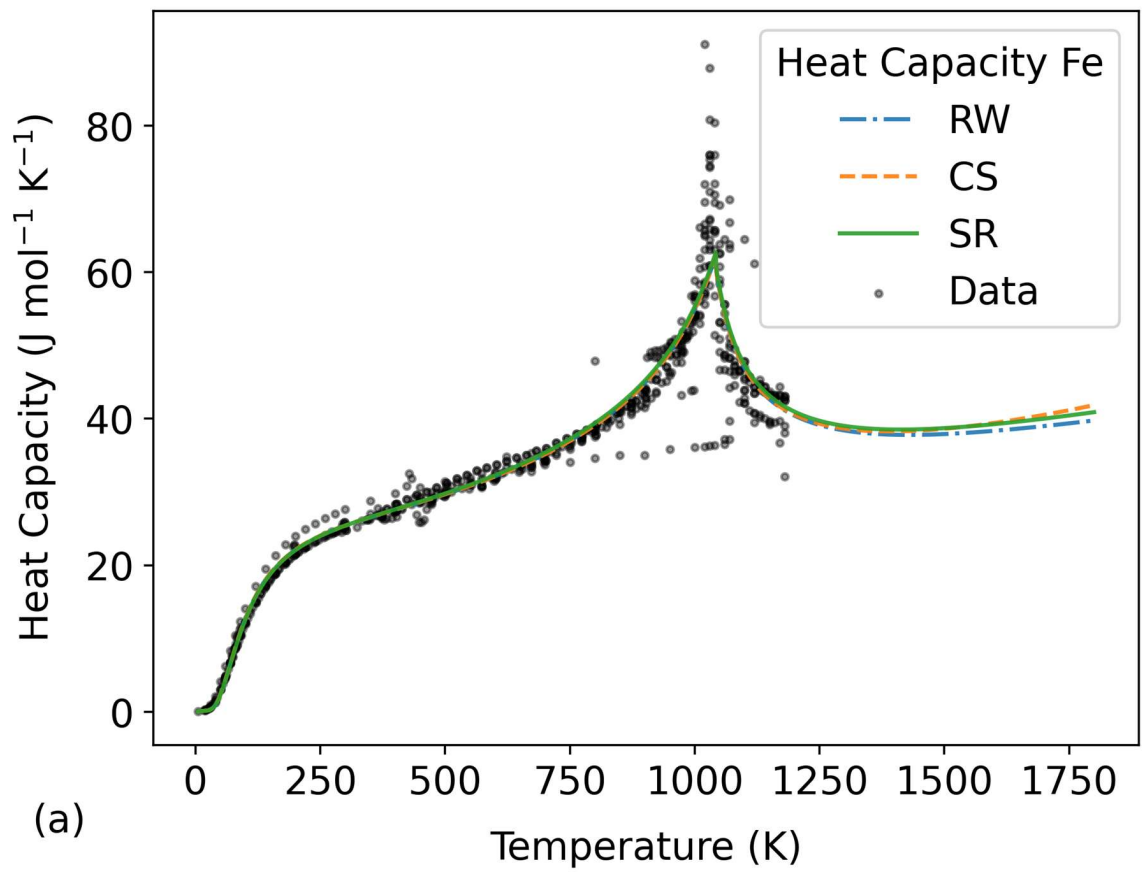

(a)

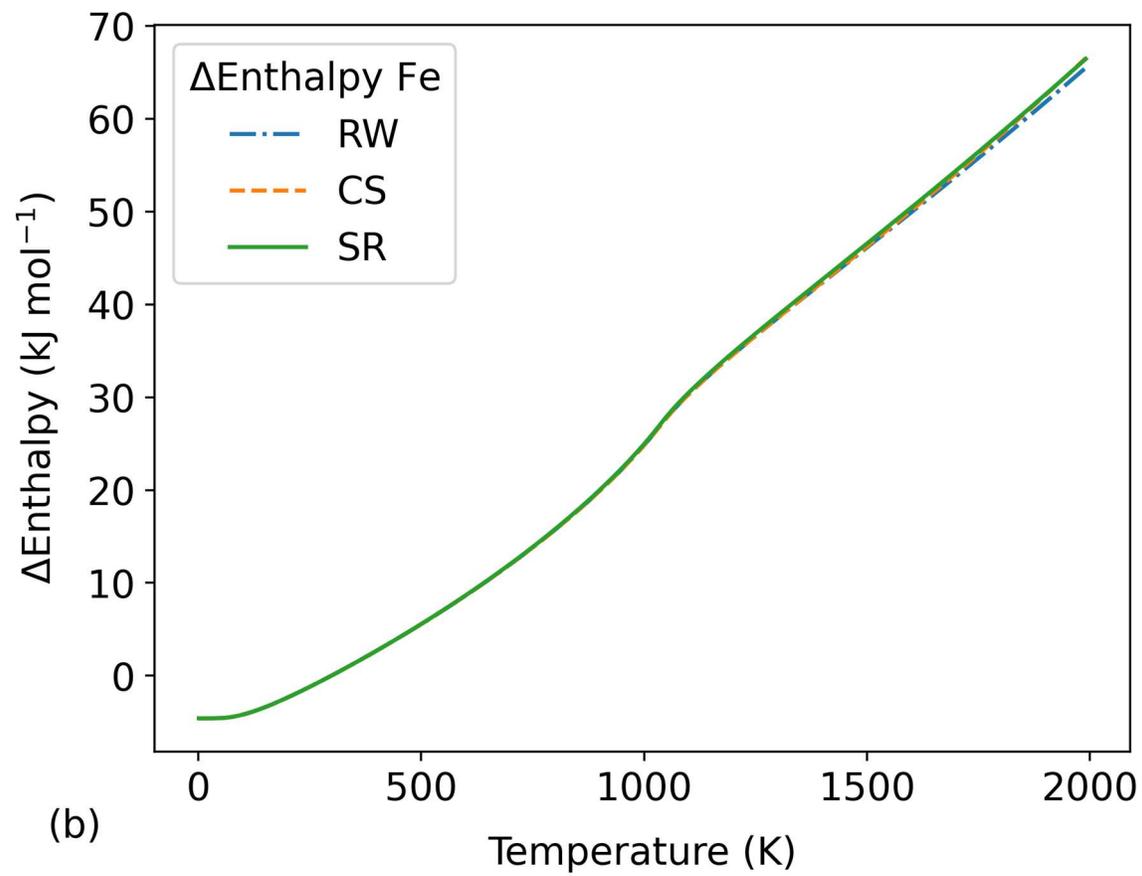

(b)

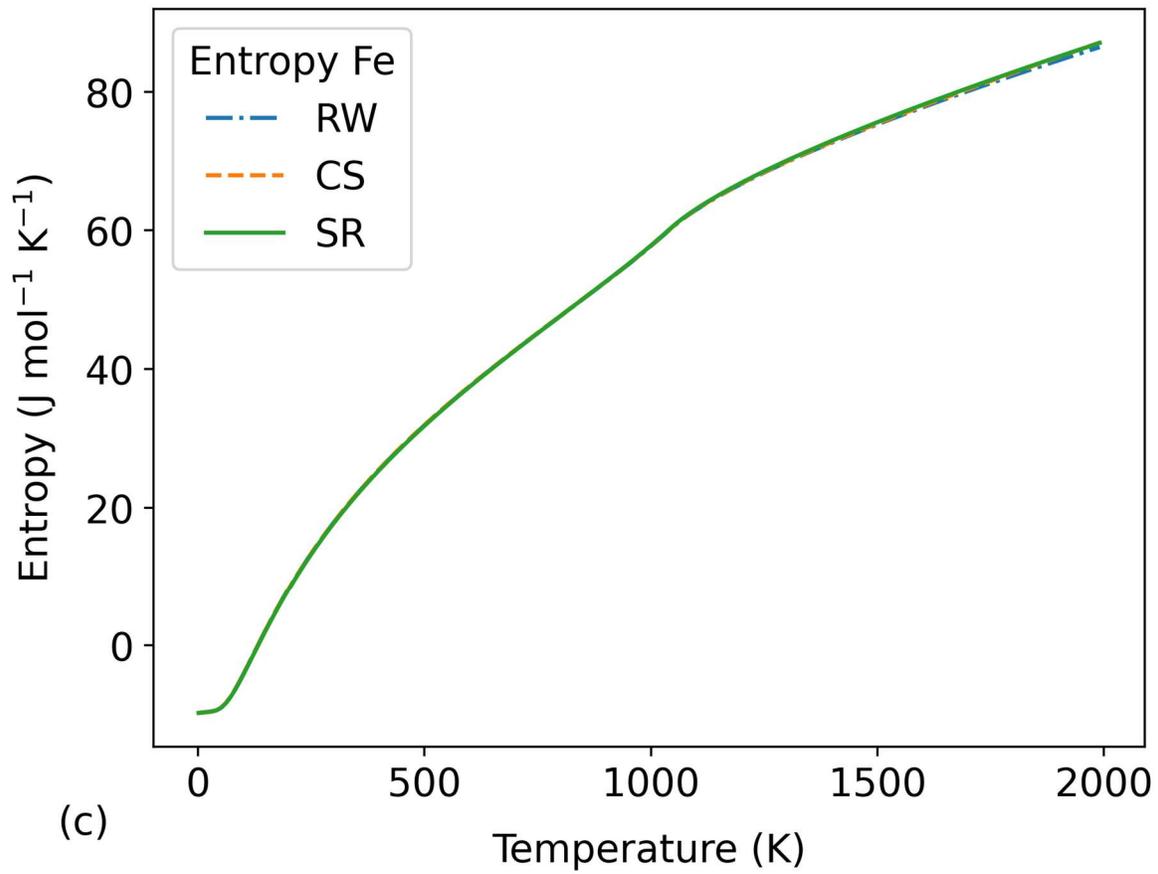

(c)

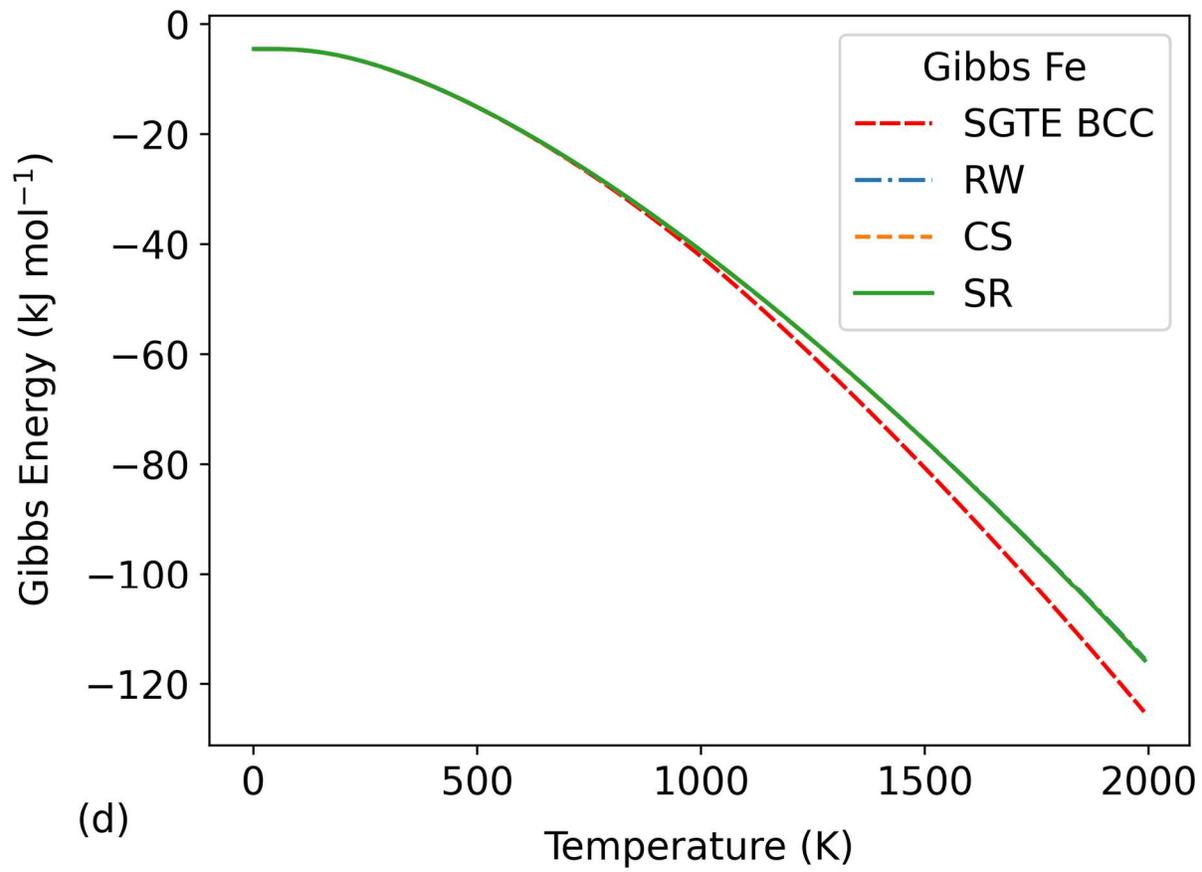

(d)

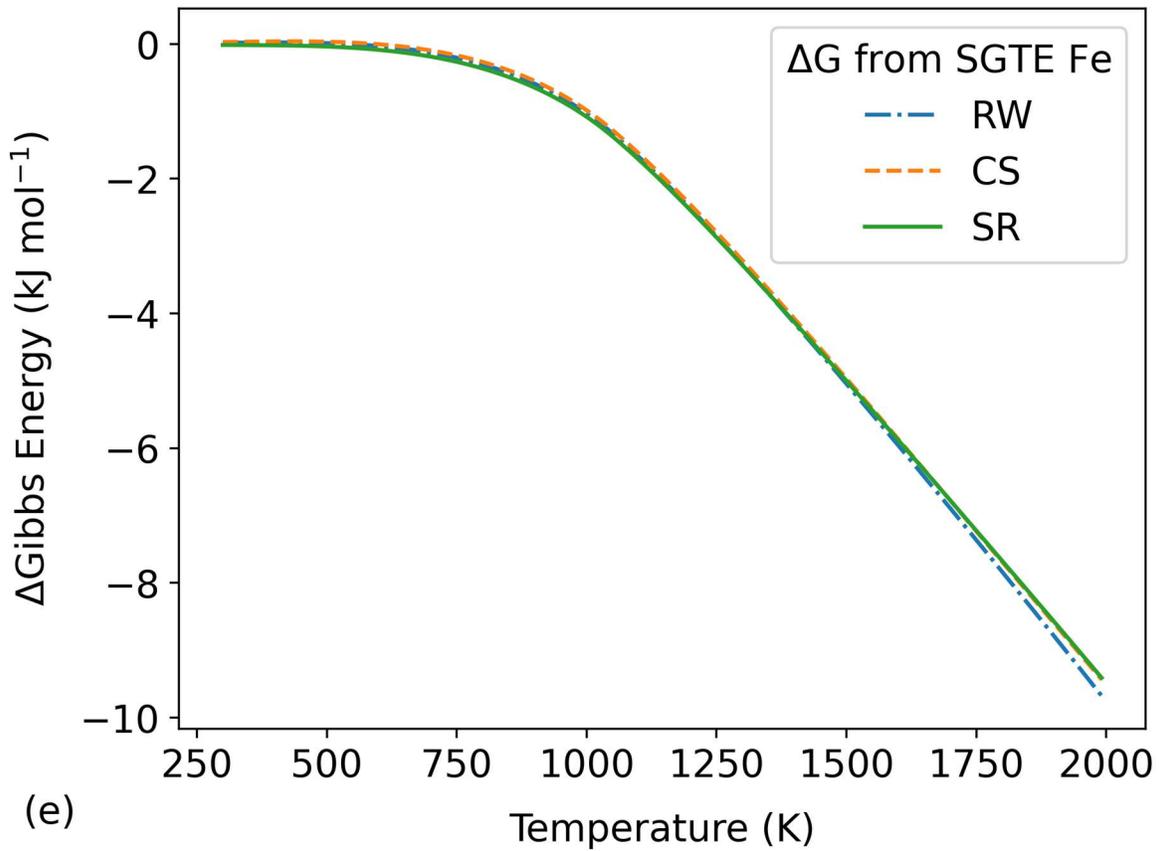

(e)

*Figure 6 Thermodynamic properties of Fe with three models: (a) Heat capacity with experimental data superimposed; (b) ΔEnthalpy, (c) Entropy, (d) Gibbs energy, and (e) Gibbs energy difference with respect to the SGTE91 database*

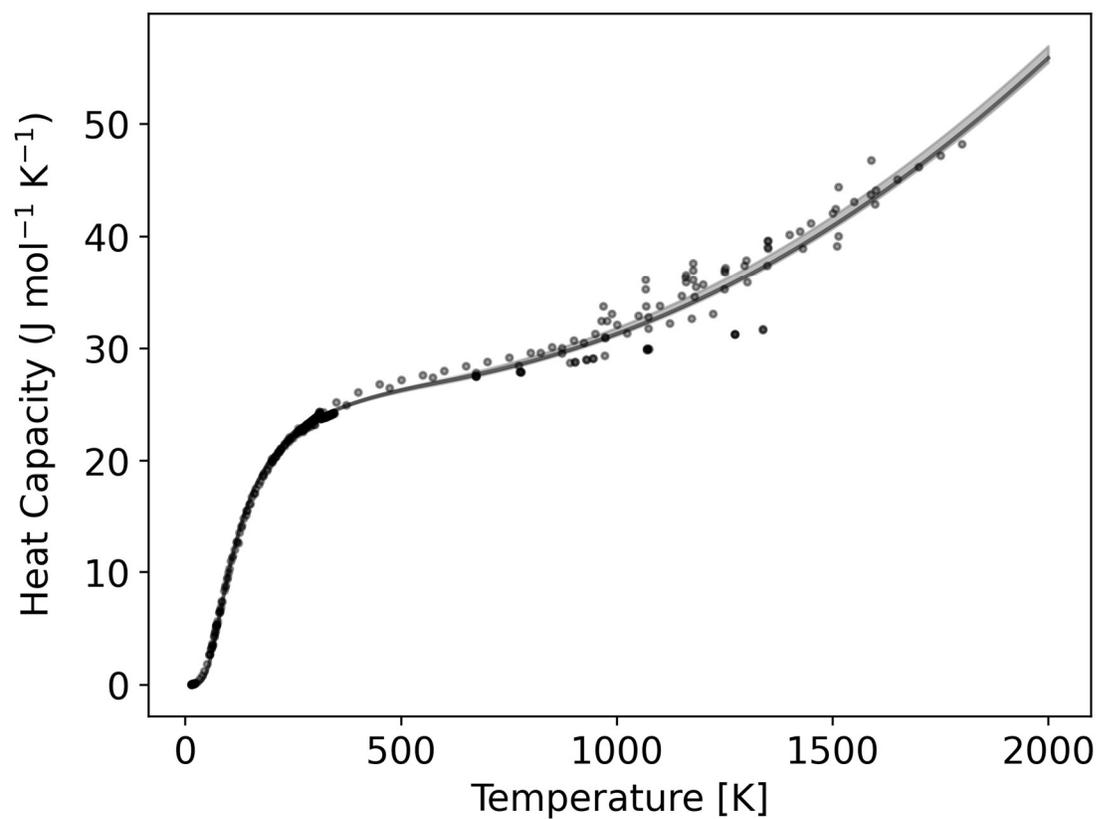

*Figure 7 Heat capacity of BCC Cr from the SR model fit by the MCMC simulation with the black line for the 50th percentile and the grey region for the 16th to 84th percentile of parameter distributions.*

*Table 1 Einstein parameter θ_E in kelvin evaluated for three models and compared to θ_E*

*recalculated from $\theta_E \cong 0.77\theta_D$ [56] collected from Kittel [66]*

| Element | RW Model | CS Model | SR Model | Kittel |
|---------|----------|----------|----------|--------|
| Ag | 152.663 | 154.105 | 157.126 | 161.220 |
| Al | 288.481 | 294.657 | 293.643 | 306.675 |
| As | 216.932 | 216.045 | 216.093 | 202.062 |
| Au | 119.064 | 119.984 | 118.060 | 118.228 |
| Be | 693.977 | 692.600 | 691.488 | 1031.805 |
| Bi | 79.841 | 79.786 | 80.281 | 85.267 |
| Ca | 167.673 | 168.106 | 168.601 | 164.802 |
| Cd | 105.736 | 106.342 | 105.181 | 149.755 |
| Co | 276.916 | 276.925 | 310.849 | 318.856 |
| Cr | 337.775 | 357.353 | 356.820 | 451.415 |
| Cs | 32.696 | 33.159 | 33.567 | 27.228 |
| Cu | 233.121 | 235.063 | 235.628 | 252.936 |
| Fe | 302.647 | 305.513 | 308.012 | 336.770 |
| Ge | 252.417 | 251.854 | 251.884 | 267.983 |
| Hf | 148.562 | 148.734 | 151.373 | 180.566 |
| Hg | 67.104 | 66.658 | 68.912 | 51.519 |
| Ir | 213.588 | 213.811 | 214.048 | 300.943 |
| K | 70.806 | 71.599 | 73.002 | 65.204 |
| La | 86.838 | 86.627 | 87.366 | 101.747 |
| Mg | 238.898 | 243.708 | 236.739 | 286.613 |
| Mn | 303.819 | 304.932 | 311.649 | 293.778 |
| Mo | 273.099 | 287.375 | 286.843 | 322.439 |
| Na | 119.313 | 120.402 | 118.284 | 113.212 |
| Nb | 202.714 | 209.529 | 203.953 | 197.046 |
| Ni | 275.288 | 282.830 | 272.713 | 322.439 |
| Pb | 62.947 | 63.204 | 65.597 | 75.236 |
| Pd | 202.244 | 203.136 | 210.445 | 196.330 |
| Pt | 172.192 | 173.854 | 173.836 | 171.968 |
| Rb | 4.551 | 46.211 | 47.278 | 40.126 |
| Re | 194.074 | 196.270 | 198.906 | 308.108 |
| Rh | 253.896 | 255.783 | 256.389 | 343.935 |
| Sb | 134.463 | 135.949 | 136.153 | 151.188 |
| Sc | 224.294 | 231.820 | 230.018 | 257.951 |
| Si | 456.413 | 457.542 | 454.617 | 462.163 |
| Sn | 121.411 | 123.778 | 121.459 | 143.306 |
| Ta | 163.756 | 166.897 | 166.724 | 171.968 |
| Tl | 67.302 | 67.451 | 67.537 | 56.248 |
| V | 273.551 | 279.449 | 281.809 | 272.282 |
| W | 229.228 | 231.816 | 235.770 | 286.613 |
| Y | 146.363 | 147.039 | 154.774 | 200.629 |
| Zn | 158.841 | 160.560 | 157.789 | 234.306 |

*Table 2 AICc values of RW, CS and SR models*

| Element | RW Model | CS Model | SR Model |
|---|---|---|---|
| Ag | 389.599 | 415.703 | **350.600** |
| Al | **251.753** | 269.512 | 261.026 |
| As | 15.597 | **15.272** | 22.461 |
| Au | 804.237 | 871.017 | **730.296** |
| Be | **18.698** | 18.708 | 22.356 |
| Bi | **577.023** | 579.292 | 642.437 |
| Ca | **51.090** | 51.179 | 54.630 |
| Cd | 223.538 | **221.784** | 348.205 |
| Co | 7116.245 | 7116.098 | **7065.640** |
| Cr | 430.023 | 399.499 | **377.888** |
| Cs | 309.743 | **301.347** | 319.562 |
| Cu | **775.554** | 798.968 | 791.484 |
| Fe | 14907.685 | 14834.926 | **14734.233** |
| Ge | 1193.508 | 1202.554 | **1172.808** |
| Hf | 128.947 | **128.803** | 142.706 |
| Hg | 449.682 | **443.437** | 526.792 |
| Ir | 57.778 | **57.503** | 90.183 |
| K | 282.080 | 208.341 | **194.546** |
| La | 36.960 | **36.433** | 40.842 |
| Mg | 132.690 | 145.240 | **124.597** |
| Mn | 105.495 | **105.270** | 142.826 |
| Mo | 1342.166 | **962.293** | 1077.956 |
| Na | 146.640 | 133.108 | **133.064** |
| Nb | **193.707** | 194.110 | 197.240 |
| Ni | 1293.893 | 1392.394 | **1225.163** |
| Pb | 229.296 | 240.192 | **186.585** |
| Pd | 119.394 | **111.726** | 286.532 |
| Pt | 619.980 | **611.384** | 616.560 |
| Rb | **56.401** | 58.036 | 130.098 |
| Re | 463.830 | 272.504 | **139.323** |
| Rh | **27.017** | 29.455 | 34.991 |
| Sb | 85.497 | **80.036** | 138.488 |
| Sc | 59.882 | 58.318 | **53.993** |
| Si | 718.435 | 732.982 | **688.499** |
| Sn | **784.704** | 823.040 | 803.525 |
| Ta | 3283.965 | **3211.131** | 3212.759 |
| Tl | **54.073** | 54.513 | 57.799 |
| V | 519.204 | **413.700** | 416.318 |
| W | 18891.667 | 13742.734 | **13332.490** |
| Y | 50.128 | 50.151 | **48.601** |
| Zn | 429.296 | 430.440 | **412.778** |